\documentclass[]{article}
\usepackage{amsmath}
\usepackage{amssymb}
\usepackage{authblk}
\usepackage{caption}
\usepackage{cite}
\usepackage{xcolor}
\usepackage{color}
\usepackage{float}
\usepackage{geometry}
\usepackage{graphicx}
\usepackage{hyperref}
\usepackage{textcomp}
\usepackage{epstopdf}
\usepackage{subcaption}
\usepackage{lscape}
\usepackage{breqn}
\newcommand\hl[1]{%
	\bgroup
	\hskip0pt\color{red!80!black}%
	#1%
	\egroup
}
\title{Analysis of scalar fields with series convolution}

\author[ ]{Emir Baysazan}
\author[ ]{Tolga Birkandan\thanks{Corresponding author. E-mail: birkandant@itu.edu.tr}}
\author[ ]{\.{I}smail Ey\"{u}phan \"{U}nver}
\affil[ ]{Department of Physics, Istanbul Technical University, 34469 Istanbul, Türkiye.}

\begin{document}
\maketitle
\begin{abstract}
\noindent Wave equations for some curved spacetimes may involve functions that prevent a solution in a closed form. In some cases, these functions can be eliminated by transformations and the solutions can be found analytically. In the cases where such transformations are not available, the infinite series expansions of these functions can be convoluted with the power series solution ansatz. We study such an example where the solution is based on a special function.
\end{abstract}
%\begin{description}
%\item[Keywords: Series convolution, Wave equations, Scalar field, Curved spacetime] 
%\end{description}
%%%%%%%%%%%%%%%%%%%%%%%%%%%%%%%%%%%%%%%%%%%%%%%%%%%%%%%%%%%%%%%%%%%%%%%%%%%%%%%%%%%%%
\section{Introduction}
Exact solutions provide a comprehensive understanding of the structure's behavior. Not only can they supply an analytical solution to the problem, but they can also provide information about the underlying symmetries such as conformal symmetry \cite{Birkandan:2011fr, Guica:2008mu}. However, exact solutions are not possible or easy to obtain for many physical systems. The reason can be either the nature of the problem or the mathematical difficulties arising from the selection of the framework. The latter obstacle can usually be solved by changing the approach to the solution (e.g. by transforming the variables). Otherwise, numerical analysis is the only possible method to solve the problem.

If the problem reduces to certain second-order differential equations, a closed-form solution in terms of well-known hypergeometric or related functions makes the analysis of the problem straightforward as the mathematical tools like the special forms of the functions or the connection of the solutions around different singular points have been studied extensively in the literature \cite{mathews, Slavyanov}.

The solutions of the Heun-type equations form another class of special functions in the analysis of second-order differential equations with polynomial coefficients \cite{Slavyanov, Ronveaux}. They are found in many applications in physics, yet their mathematical tools are not as developed as the hypergeometric functions \cite{Hortacsuheun}. Therefore the power series solutions of the Heun-type equations are inevitable for obtaining physical results from the systems having those functions as solutions \cite{Vieira:2016ubt, Motohashi:2021zyv, Hortacsu:2021dly, Hatsuda:2020sbn, Hortacsu:2020bee, Willenborg:2023ixu}. The general Heun function has a three-term recurrence relation whereas the double-confluent Heun (DCH) function has a seven-term recurrence relation \cite{Birkandan:2020clt, maple}. 

In physical problems, we mainly encounter the general and singly-confluent Heun equations. There are fewer applications in which the DCH equation emerges. Dariescu group solved the Dirac equation in terms of these functions in several cases \cite{Dariescu:2012gb, Dariescu:2021zve}. Vieira \cite{Vieira:2015uua} and Bezerra et al. \cite{Bezerra:2013iha} solved the Klein-Gordon equation in curved spacetimes in terms of DCH functions. We have solved the Klein-Gordon and Dirac equations in terms of the DCH function in our previous works based on Nutku's helicoid solution \cite{Aliev:1998cu, Nutku:1996dn, Aliev:1996sf, Aliev:1998xu, Birkandan:2006ac, Birkandan:2007ey, Birkandan:2007cw}. 

In a recent work based on Nutku's solution with a non-trivial addition of the time coordinate, we also encountered this function as the solution of the Klein-Gordon equation \cite{Birkandan:2022zny}. However, for the full radial equation, we came across an equation that has the same singularity structure as the DCH equation but with non-polynomial coefficients. In \cite{Birkandan:2022zny}, we have called this equation the ``deformed" double-confluent Heun equation (DDCH) and studied it numerically.

In the present paper, we will study the series solution of the ``deformed" double-confluent Heun equation by employing the Cauchy product for the convolution of the series expansions of the functions present in the coefficients with the power series solution ansatz of the equation. The Cauchy product \cite{bloch} is an important tool for convolution and it has applications in gravity \cite{Gurtug:2023pfp, Beachley:2018tux, DelPiano:2024gvw, Schwartz:2018pnh, Harko:2016wwn}. 

All computational applications in this work are performed on SageMath computer algebra system \cite{sagemath}.

In the next section, we introduce the Nutku-Ghezelbash-Kumar metric and the Klein-Gordon equation for a massive and charged scalar field. In the third section, we study the solutions of the double-confluent Heun equation which emerges from the angular equation and the $c_2=0$ limit of the radial equation. The fourth section is dedicated to the study of the full radial equation which is defined as a ``deformed" double-confluent Heun equation. This section is based on the convolution of the $tanh^{-1}$ and solution series. We give our last comments in the last section. Appendix A involves the coefficient matrices and series coefficients of the solutions used in the text. In Appendices B and C, we present a graphical analysis of the series solutions of the DCH and DDCH equations for various parameters.
%%%%%%%%%%%%%%%%%%%%%%%%%%%%%%%%%%%%%%
\section{The NGK Metric}
The Nutku-Ghezelbash-Kumar (NGK) metric is a five-dimensional solution based on Nutku's instanton metric for helicoid/catenoid minimal surface \cite{Ghezelbash:2017bjs}. It is a solution of Einstein-Maxwell equations and the metric reads
\begin{align}\label{metrikk}
ds^2=&-\frac{dt^2}{(H(r))^2}+\frac{H(r)}{\sqrt{1+\frac{a^2}{r^2}}}\bigg[dr^2+(a^2+r^2)d\theta^2 +\bigg(1+\frac{a^2sin^2\theta}{r^2} \bigg)dy^2+ \bigg(1+\frac{a^2cos^2\theta}{r^2} \bigg)dz^2-\frac{a^2sin(2\theta)dydz}{r^2}\bigg],
\end{align}
where $r \in (0,\infty)$, $\theta \in [0,2\pi]$, and the Killing coordinates $y$ and $z$ are periodic on a 2-torus \cite{Ghezelbash:2017bjs}.
For the helicoid case, we have
\begin{equation}
H(r)=1+a c_2 sinh^{-1} \bigg( \frac{r}{a} \bigg),
\end{equation}
which is related with the electromagnetic (EM) potential. The only non-zero component of the EM potential is
\begin{equation}
A_t=\sqrt{\frac{3}{2}} \frac{1}{H(r)},
\end{equation}
as given in \cite{Ghezelbash:2017bjs}. $a \geq 0$ and $c_2 \geq 0$ are required to prevent a signature change in the metric. 

The Klein-Gordon equation for a massive (with mass $\mu$) and charged (with charge $q$) scalar field
\begin{equation}
[(\nabla ^\nu-iqA^\nu)(\nabla_\nu-iqA_\nu)-\mu^2]\Phi(t,r,\theta,y,z)=0,    
\end{equation}
in the background of the NGK metric is studied in \cite{Birkandan:2022zny} and its angular and radial parts are given as
\begin{align}
&\frac{d^2R(r)}{dr^2}+\frac{r}{a^2+r^2}\frac{dR(r)}{dr}+\frac{r\sqrt{a^2+r^2}H(r) \bigg[\sqrt{6}qH(r)\omega+(H(r))^2\omega^2+\frac{3}{2}q^2-\mu^2 \bigg]-(m_1^2+m_2^2)r^2-\lambda}{a^2+r^2} R(r)=0, \label{radial}\\
&\frac{d^2 S(\theta)}{d\theta^2}-\bigg[a^2(m_1 cos \theta+m_2 sin \theta)^2 -\lambda\bigg]S(\theta)=0, \label{angulareqn}
\end{align}
for the ansatz $\Phi(t,r,\theta,y,z)=e^{-i \omega t} e^{i m_1 y} e^{i m_2 z} R(r)S(\theta)$ of the scalar field.

The angular equation and the $c_2=0$ limit of the radial equation can be written in the form of the DCH equation. The full radial equation ($c_2 \neq 0$) is called the ``deformed" double-confluent Heun equation as its structure and numerical behavior resemble the DCH equation \cite{Birkandan:2022zny}.

It should be noted that, we performed the transformations $r = asinh(x)$ and $x = tanh^{-1}(u)$, and worked with the radial coordinate $u$ with range $(0,1)$  \cite{Birkandan:2022zny}.
%%%%%%%%%%%%%%%%%%%%%%%%%%%%%%%%%%%%%%%%%%%%%%%%%%%%%%%%%%%%%%%%%%%%%%%%
\section{General form of the double-confluent Heun equation}
The general form of the DCH equation reads
\begin{equation} \label{heund}
	\frac{d^2 H_D}{du^2}
	+{\frac {2\,{u}^{5} -\alpha\,{u}^{4}-4\,{u}^{3}+2\,u+\alpha}{ \left( u-
			1 \right) ^{3} \left( u+1 \right) ^{3}}}
	\frac{dH_D}{du}
	+{\frac {\beta{u}^{2}+ \left( \gamma+2\,\alpha \right) u+\delta}{
			\left( u-1 \right) ^{3} \left( u+1 \right) ^{3}}}
	H_D=0,
\end{equation}
and the irregular, rank-1 singular points are located at $u=\{-1, 1\}$. For a detailed analysis of the equation, one can see \cite{Ronveaux,Slavyanov,Birkandan:2020clt}.

We can construct a series solution around $u=0$ as
\begin{equation}\label{series}
	H_D(u)=\sum_{n=0}^{\infty}b_n u^n.
\end{equation}
Applying the series (\ref{series}) in (\ref{heund}), we obtain the recurrence relation
\begin{align}\label{recur}
&n(n+1) b_{n}-\alpha(n +1) b_{n+1}-[(n+2)(3n+7)-\beta] b_{n+2}+(2 \alpha +\gamma ) b_{n+3} \nonumber \\
&+[(n+4)(3n+11)+\delta] b_{n+4}+\alpha(n +5) b_{n+5}-(n + 5)(n + 6) b_{n+6}=0.
\end{align}
In the literature, most of the applications of the DCH equation employ the definition used in the Maple computer algebra system \cite{maple}. Accordingly, the values of the first two coefficients of the DCH series solution will be set to $b_0=0$, $b_1=1$. We also need $b_2, b_3, b_4$ and $b_5$ in order to work with the 7-term recurrence relation. These coefficients are read from the series solution as
\begin{align}
b_2&=\frac{\delta}{2},\\
b_3&=\frac{(2+\delta)\alpha+\gamma}{6}, \\
b_4&=\frac{(2+\delta)\alpha^2+\alpha\gamma+\delta^2+10\delta+2\beta}{24}, \\
b_5&=\frac{(2+\delta)\alpha^3+\alpha^2 \gamma+2\alpha (\delta^2+21\delta+\beta+24)+4\gamma(\delta+6)}{120}.
\end{align}
Using the recurrence relation, we can write a matrix equation for the coefficients. For example, for 15 terms, the equation is
\begin{equation}
	M_{DCH}\left(\begin{array}{r}
		b_2 \\
		b_3 \\
		\vdots \\
		b_{14}
	\end{array}\right)=v_{DCH},
\end{equation}
where $M_{DCH}$ and $v_{DCH}$ are given by equations (\ref{coeffmatrixdch}) and (\ref{coeffvectors}), respectively. For $\alpha=0$ we get the coefficients given by the equations (\ref{DCHcoeffsini}-\ref{DCHcoeffsfin}). One should note that for $\alpha=0$, the seven-term recurrence relation reduces to a five-term relation.

\begin{figure}[H]
	\begin{subfigure}{.5\textwidth}
		\centering
		% include first image
		\includegraphics[width=1.0\linewidth]{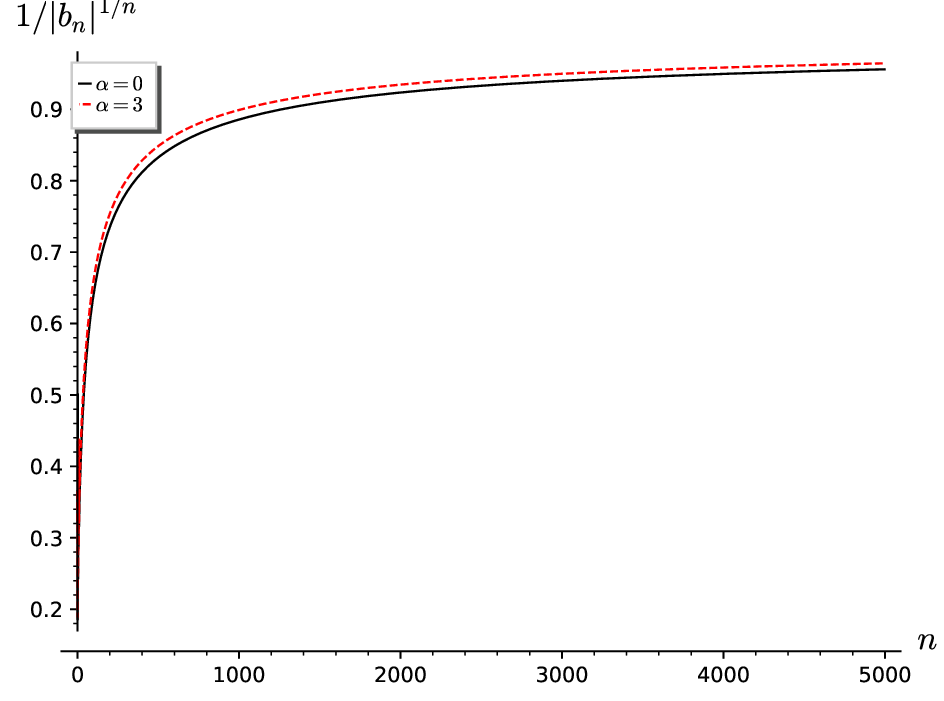}
		\caption{Varying $\alpha$}
		\label{fig:DCH_Series_1_a0}
	\end{subfigure}
	\begin{subfigure}{.5\textwidth}
		\centering
		% include first image
		\includegraphics[width=1.0\linewidth]{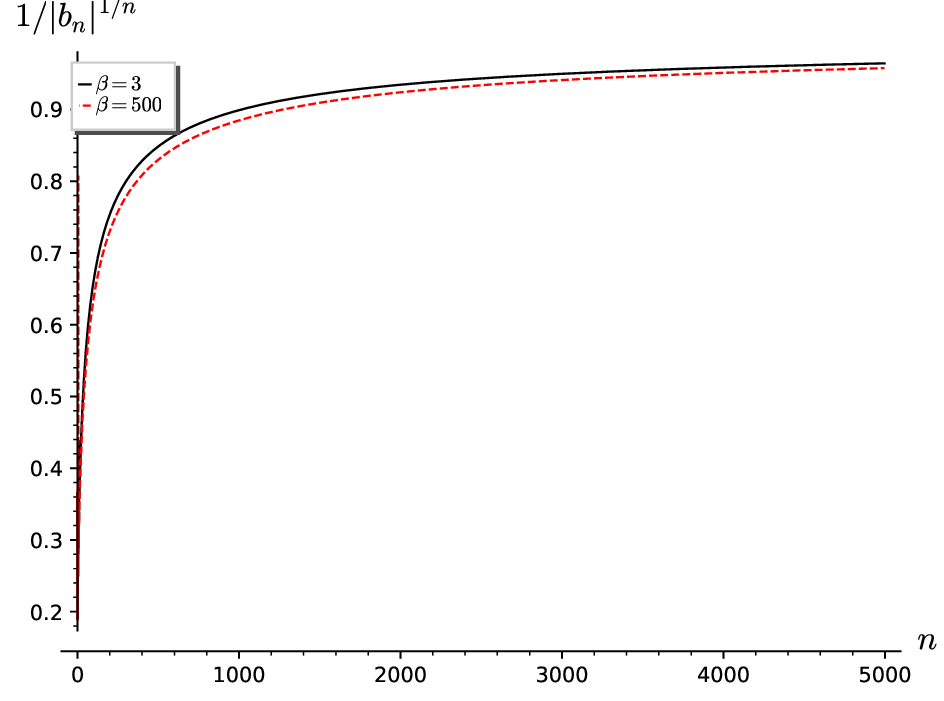}
		\caption{Varying $\beta$}
		\label{fig:DCH_Series_1}
	\end{subfigure}
    \begin{subfigure}{.5\textwidth}
		\centering
		% include second image
		\includegraphics[width=1.0\linewidth]{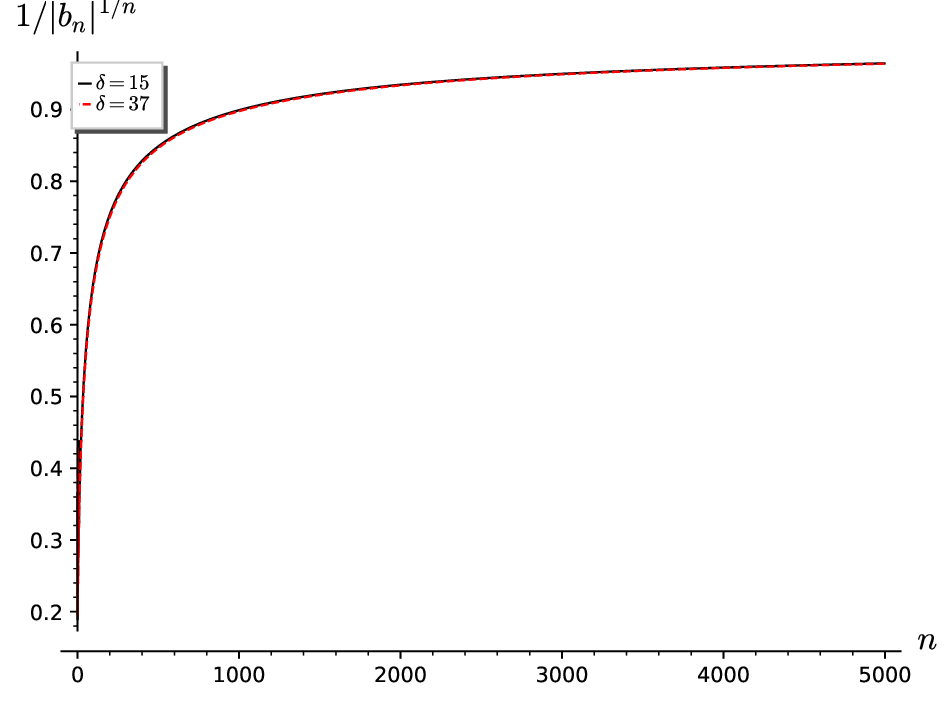}
		\caption{Varying $\delta$}
		\label{fig:DCH_Series_2_a0}
	\end{subfigure}
    \begin{subfigure}{.5\textwidth}
		\centering
		% include second image
		\includegraphics[width=1.0\linewidth]{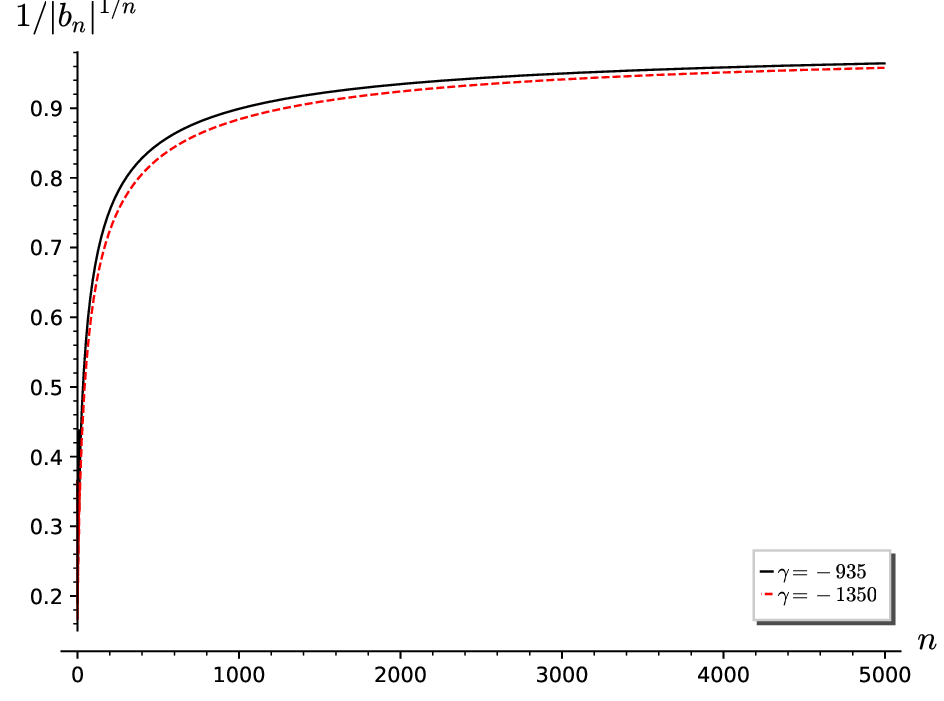}
		\caption{Varying $\gamma$}
		\label{fig:DCH_Series_2}
	\end{subfigure}
	\caption{The evolution of $\frac{1}{|b_n|^{1/n}}$ for the radius of convergence. The numerical values of the parameters are \{$\alpha=3, \beta=3, \gamma=-935, \delta=15 $\} if not stated otherwise.}
	\label{fig:radconvdch}
\end{figure}

The radius of convergence of the series solution can be determined from the theorem given in \cite{gamelin} on the root test. The theorem states that if $|b_n|^{1/n}$ has a limit as $n \rightarrow \infty$, either finite or $+\infty$, then the radius of convergence $R$ of the series (\ref{series}) is found by
\begin{equation}
R=\frac{1}{\lim_{n\to\infty} (|b_n|^{1/n})}.
\end{equation}
In Figure (\ref{fig:radconvdch}), it is seen that as $n$ grows larger, the radius convergence $R \rightarrow 0.9 \sim 1$ and the behavior is independent of the parameter variations for large $n$.

\begin{figure}[H]
	\begin{subfigure}{.5\textwidth}
		\centering
		% include first image
		\includegraphics[width=1.0\linewidth]{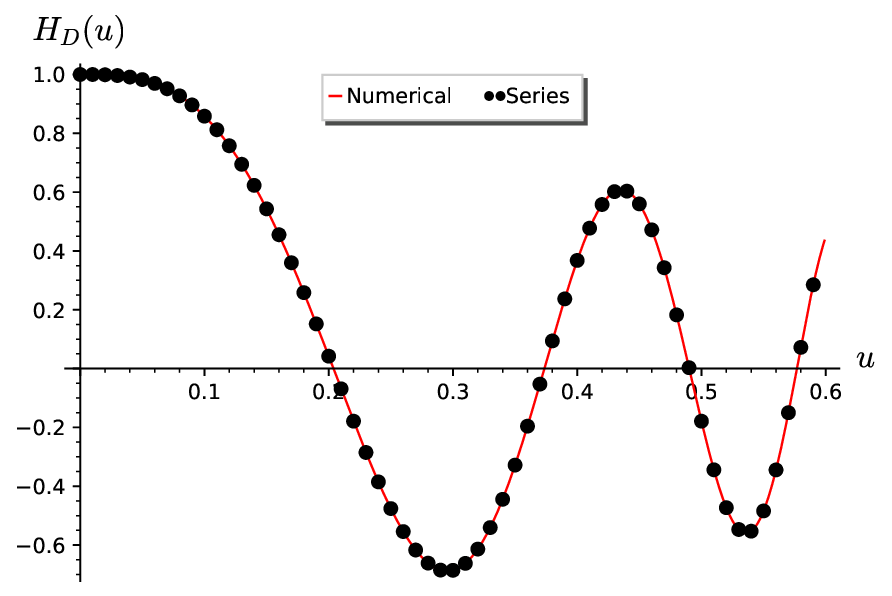}
		\caption{$\alpha=0$.}
		\label{fig:sub-firstang0}
	\end{subfigure}
	\begin{subfigure}{.5\textwidth}
		\centering
		% include second image
		\includegraphics[width=1.0\linewidth]{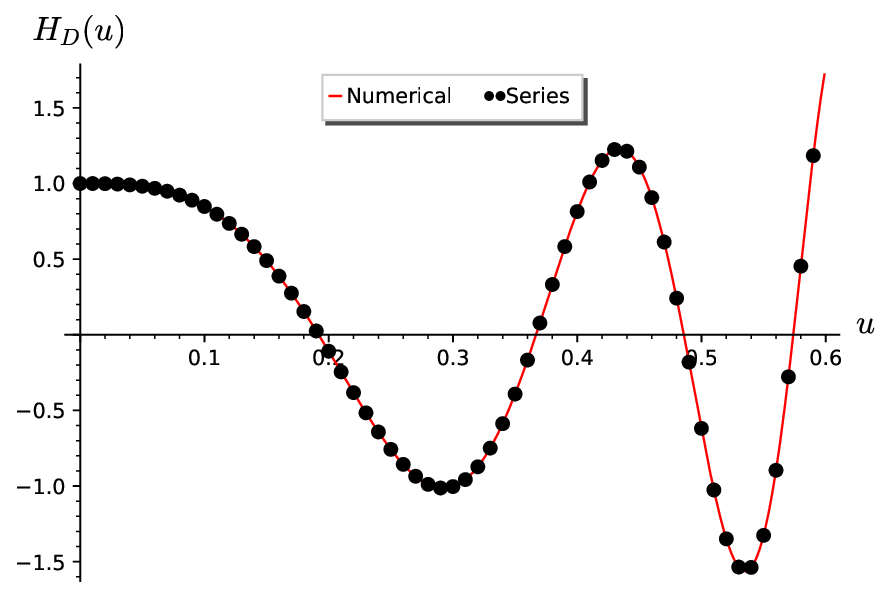}
		\caption{$\alpha=3$.}
		\label{fig:sub-secondang0}
	\end{subfigure}
	\caption{Numerical solution and the series solution (with $100$ terms) of the DCH equation (\ref{heund}) with \{$ \beta=2.06, \gamma=-895.59, \delta=1.0 $\}.}
	\label{fig:DCHnumerical}
\end{figure}

Moreover, the radius of convergence of the power series solution of the equation (\ref{heund}) can be at most the distance to the closest singular point of the equation which is located at $u=-1$, and $u=1$ \cite{kris}. Thus we conclude that the radius of convergence of the local solution $H_D(\alpha,\beta,\gamma,\delta,u)$ given by the series (\ref{series}) around the ordinary point $u=0$ is $R=1$ and consequently $|u|<1$. Another theorem states that the power series $\sum_{n=0}^{\infty}b_n u^n$ converges absolutely if $|u|<R$ and it is also known that every absolutely convergent series is convergent \cite{gamelin}.
Figure (\ref{fig:DCHnumerical}) shows the agreement of the series solution with the numerical solution for zero and non-zero $\alpha$ values. 

For the NGK metric, the angular part and the $c_2=0$ limit of the radial part of the Klein-Gordon equation can be written in terms of the DCH equation with $\alpha=0$ \cite{Birkandan:2022zny}. However, for $c_2 \neq 0$ we have the ``deformed double-confluent Heun equation" as analyzed in \cite{Birkandan:2022zny} and in the following section.
In order to see the convergence of the DCH series solution, we will analyze the behavior of the partial sums of the series, namely $S_N=\sum_{n=0}^{N}b_n u^n$ and take the limit $S=\lim_{N\to\infty} S_N$ to see if they converge to a finite value \cite{mathews, bloch}. The partial sums \{$S_N$\} constitute a sequence and \{$S_N$\} is a Cauchy sequence if for each $\epsilon>0$ there exists an integer $N$ such that $n, m>N$ with $|S_n - S_m|<\epsilon$ \cite{bak}. In other words, for a Cauchy sequence, $|S_n - S_m| \rightarrow 0$ as $n, m \rightarrow \infty$ \cite{szekeres}. As stated in \cite{bak}, \{$S_N$\} converges to $S$ if and only if \{$S_N$\} is a Cauchy sequence.

\begin{figure}[H]
	\begin{subfigure}{.5\textwidth}
		\centering
		% include first image
		\includegraphics[width=1.0\linewidth]{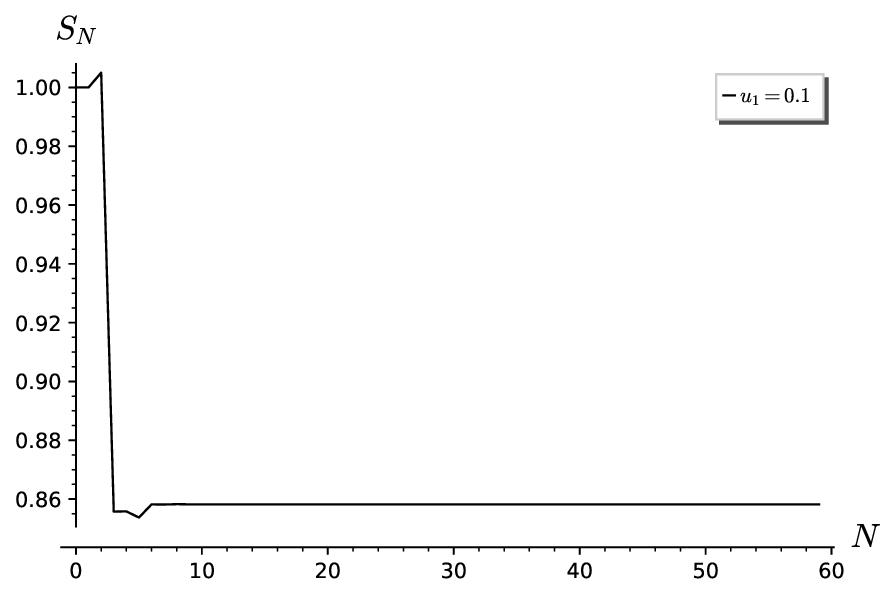}
		\caption{$\alpha=0, u=0.1$}
		\label{fig:DCH_Series_1_a0}
	\end{subfigure}
	\begin{subfigure}{.5\textwidth}
		\centering
		% include first image
		\includegraphics[width=1.0\linewidth]{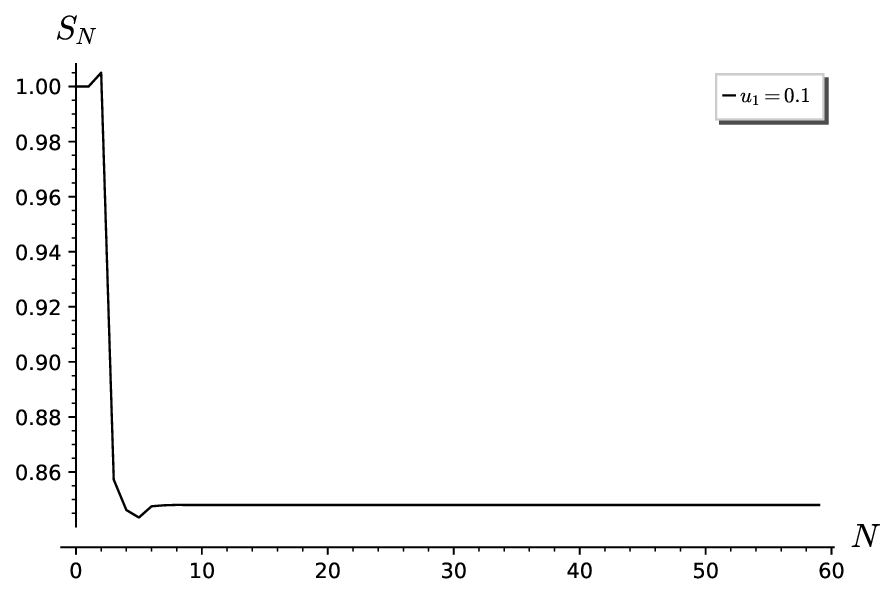}
		\caption{$\alpha=3, u=0.1$}
		\label{fig:DCH_Series_1}
	\end{subfigure}
    \begin{subfigure}{.5\textwidth}
		\centering
		% include second image
		\includegraphics[width=1.0\linewidth]{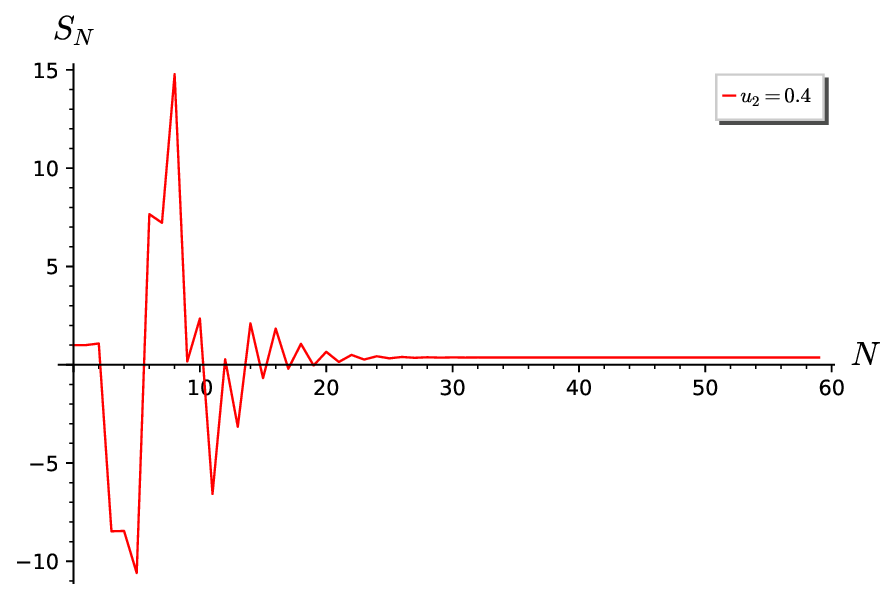}
		\caption{$\alpha=0, u=0.4$}
		\label{fig:DCH_Series_2_a0}
	\end{subfigure}
    \begin{subfigure}{.5\textwidth}
		\centering
		% include second image
		\includegraphics[width=1.0\linewidth]{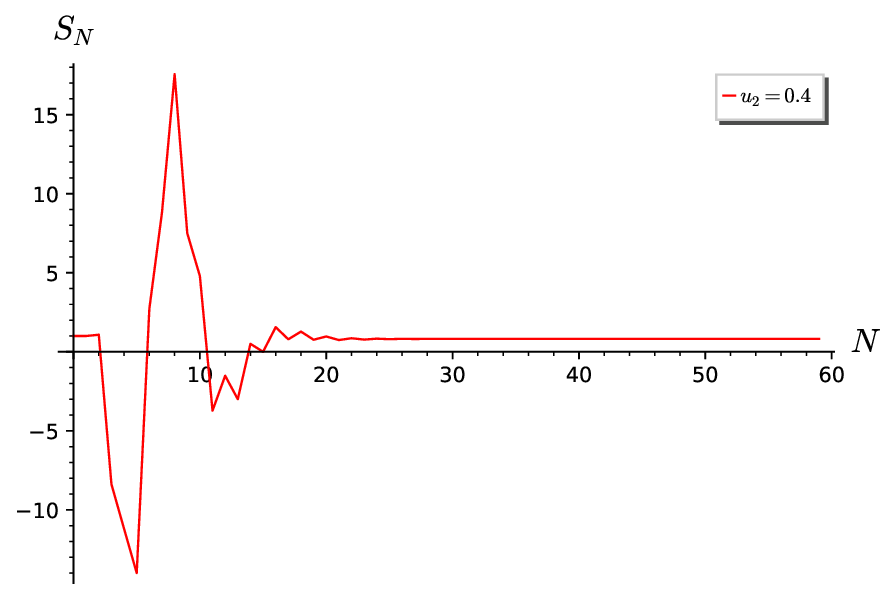}
		\caption{$\alpha=3, u=0.4$}
		\label{fig:DCH_Series_2}
	\end{subfigure}
        \begin{subfigure}{.5\textwidth}
		\centering
		% include second image
		\includegraphics[width=1.0\linewidth]{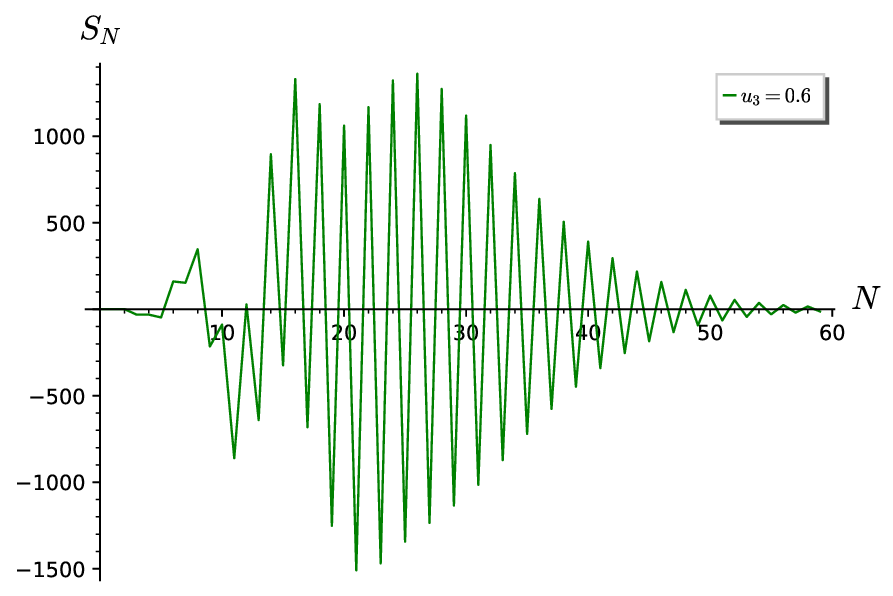}
		\caption{$\alpha=0, u=0.6$}
		\label{fig:DCH_Series_3_a0}
	\end{subfigure}
        \begin{subfigure}{.5\textwidth}
		\centering
		% include second image
		\includegraphics[width=1.0\linewidth]{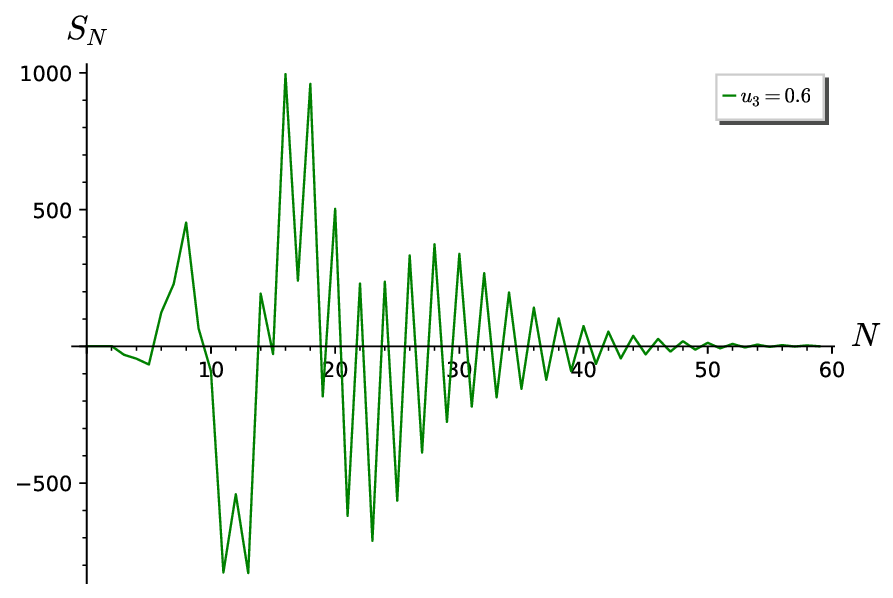}
		\caption{$\alpha=3, u=0.6$}
		\label{fig:DCH_Series_3}
	\end{subfigure}
	\caption{$\bigg(S_N= \sum_{n=0}^{N}b_n u^n$ vs. $N$$\bigg)$. Convergence of the series solution for the DCH equation (\ref{heund}) with \{$ \beta=2.06, \gamma=-895.59, \delta=1.0 $\}.}
	\label{fig:DCHcompare}
\end{figure}

In Figures (\ref{fig:DCH_Series_1_a0},\ref{fig:DCH_Series_2_a0},\ref{fig:DCH_Series_3_a0}) for $\alpha=0$, the series solutions asymptotically converge to $S=0.85, 0.37$ and $0.45$ for $u_1=0.1, u_2=0.4$ and $u_3=0.6$, respectively. The Figures (\ref{fig:DCH_Series_1},\ref{fig:DCH_Series_2},\ref{fig:DCH_Series_3}) for $\alpha=3$ show that the solutions asymptotically converge to $S=0.85, 0.81$ and $1.77$ for $u_1=0.1, u_2=0.4$ and $u_3=0.6$, respectively. These values are compatible with the numerical solutions given in Figure (\ref{fig:DCHnumerical}). In all figures, $|S_n - S_m| \rightarrow 0$ as $n, m \rightarrow \infty$ condition is also satisfied.

We observe that, as $u$ grows larger, more terms are needed in the series for convergence. For example, for $u=0.1$, we need around $10$ terms to reach $S$. We need around $30$ terms for $u=0.4$, and around $60$ terms for $u=0.6$ to converge their $S$ values.

Appendix B involves the graphics of the solutions and their convergence behaviors for all DCH parameters. We see that the convergence behaviors are independent of the parameters (all converge to $S$ with around 30 terms for $u=0.4$ as mentioned above), and series solutions (all with $n=100$ terms) agree with the numerical solutions.
%%%%%%%%%%%%%%%%%%%%%%%%%%%%%%%%%%%%%%%%%%%%%%%%%%%%%%%%%%%%%%%%%%%%%%%%%%%%%%%%%%%%%
\section{Deformed double-confluent Heun equation}
We will define the ``deformed" double-confluent Heun (DDCH) equation as an equation that has the same singularity structure as the DCH equation but with non-polynomial coefficients that necessitate a convolution approach for series solutions. Such an example arises from the radial part of the Klein-Gordon equation for a massive and charged scalar field in the background of the NGK metric as given by the equation (\ref{radial}). Here, $H(r)=1+a c_2 sinh^{-1} (\frac{r}{a})$ is related with the electromagnetic potential of the metric (\ref{metrikk}). 

The change of variable $r = asinh(x)$ in equation (\ref{radial}) yields
\begin{equation}\label{yardimci}
\frac{d^2 R(x)}{dx^2}+\mathcal{F}(x)R(x)=0,
\end{equation}
where
\begin{align}
\mathcal{F}(x)=&\frac{a^2}{2}(m_1^2+m_2^2)-\lambda-\frac{a^2}{2}(m_1^2+m_2^2)cosh(2x) +\frac{a^2}{2} \bigg( \sqrt{6} q \omega (H(x))^2 +  \omega^2 (H(x))^2  +\frac{3}{2}q^2 - \mu^2  \bigg) sinh(2x).
\end{align}
Here, $H(x)=1+a c_2 x$.

The singularity structure of the equation (\ref{yardimci}) can be studied more effectively by applying the change of variable $x = tanh^{-1}(u)$. Then we get,
\begin{align}\label{inceleradial}
\frac{d^2 R(u)}{du^2}+\frac{2u}{u^2-1}\frac{dR(u)}{du}+\frac{F_1(u)+F_2(u,c_2)}{(u-1)^3 (u+1)^3}R(u)=0,
\end{align}
where
\begin{align}
&F_1(u)=[(m_1^2+m_2^2)a^2-\lambda] u^2 - a^2 \bigg(\frac{3}{2} q^2-\sqrt{6} q \omega +\omega^2 - \mu^2 \bigg) u+\lambda,\\
&F_2(u,c_2)=\bigg[-a^2 \omega^2 (tanh^{-1}(u))^2  c_2^2 - a \omega (\sqrt{6} q+3 \omega) tanh^{-1}(u)   c_2-(2 \sqrt{6} q \omega-\mu^2+\frac{3}{2} q^2+3\omega^2)  \bigg]a^3 c_2 u tanh^{-1}(u), \label{F2uc2}
\end{align}
related with the parameters of the metric and the scalar field.

The final form of the $H$ function is $H(x)=1+a c_2 tanh^{-1}(u)$ and this is the key element of the equation which introduces the $c_2$ (i.e. electromagnetic) dependence into the equation through the function $F_2(u,c_2)$. The $tanh^{-1}(u)$ terms arise from this dependence which necessitates convolution for the series solution.

The case with $c_2=0$ yields the DCH equation with
\begin{align}
\alpha&=0,\\
\beta&=(m_1^2+m_2^2)a^2-\lambda,\\
\gamma&=- a^2 \bigg(\frac{3}{2} q^2+\sqrt{6} q \omega +\omega^2 - \mu^2 \bigg),\\
\delta&=\lambda.
\end{align}

In order to present its resemblance to the DCH equation (\ref{heund}), we can rewrite the equation (\ref{inceleradial}) as
\begin{align}\label{defheund}
	\frac{d^2 R(u)}{du^2}+{\frac {2\,{u}^{5} -4\,{u}^{3}+2\,u}{ \left( u-
			1 \right) ^{3} \left( u+1 \right) ^{3}}}\frac{dR(u)}{du}+\frac{F_1(u)+F_2(u,c_2)}{(u-1)^3 (u+1)^3}R(u)=0.
\end{align}
Here,
\begin{align}
	&F_1(u)=A_1 u^2+A_2 u+A_3, \\
	&F_2(u,c_2)=[A_4 (tanh^{-1}(u))^3+A_5 (tanh^{-1}(u))^2+A_6 (tanh^{-1}(u))]u,
\end{align}
where
\begin{align}
	A_1&=(m_1^2+m_2^2)a^2-\lambda, \\
	A_2&=- a^2 \bigg(\frac{3}{2} q^2+\sqrt{6} q \omega +\omega^2 - \mu^2 \bigg), \\
	A_3&=\lambda,\\
	A_4&=-a^5 \omega^2c_2^3,\\
	A_5&=-a^4 \omega c_2^2 (\sqrt{6}q+3\omega),\\
	A_6&=-a^3 c_2 (2\sqrt{6} q \omega-\mu^2+\frac{3}{2}q^2+3\omega^2).
\end{align}
The equation (\ref{defheund}) has two irregular singular points of rank-1 located at $u = \{-1, 1\}$, which is the singularity structure of the DCH equation. We will consider $tanh^{-1}(u)$ terms as deformations to the original DCH equation as they do not change the singularity structure. Therefore we will call the equation (\ref{defheund}) ``deformed" double-confluent Heun equation (DDCH).

The equation (\ref{defheund}) with $A_4=A_5=A_6=0$ (or, equally, $c_2=0$) reduces to the equation (\ref{heund}) with $\alpha=0$. Then we read $\beta=A_1$, $\gamma=A_2$ and $\delta=A_3$ in this limit. The coefficient of $\frac{dR(u)}{du}$ in (\ref{defheund}) also forces $\alpha=0$ if we demand a DCH format.

The range of the radial coordinate $u$ is $u \in (0,1)$ \cite{Birkandan:2022zny}. Using this range, we can construct a series solution around the ordinary point $u=0$ as
\begin{equation}
R(u)=\sum_{n=0}^{\infty}b_n u^n,
\end{equation}
and for $|u|< 1$, we have \cite{nist}
\begin{equation}
tanh^{-1}(u)=\sum_{n=0}^{\infty} \frac{u^{2n+1}}{2n+1}.
\end{equation}
The series solution of Eq. (\ref{defheund}) requires the convolution of these series via the Cauchy product, namely,
\begin{equation}
\bigg(\sum_{n=0}^{\infty} T_n \bigg) \bigg(\sum_{n=0}^{\infty} P_n \bigg)=\sum_{n=0}^{\infty}\sum_{k=0}^{n}T_k P_{n-k}.
\end{equation}
where $T_n=b_n u^n$ and $P_n= \frac{u^{2n+1}}{2n+1}$ \cite{bloch}. The terms such as $(tanh^{-1}(u))^3 R(u)$ are handled by consecutive Cauchy products of the series.

Unlike the DCH case, we do not have a recurrence relation in closed form. Nevertheless, we can determine the series coefficients by solving the matrix equation. For this purpose, we will take the initial coefficients same as in the DCH equation ($b_0=0$ and $b_1=1$) to see the effect of the deformation on the original DCH equation.

For example, for 15 terms, the matrix equation is
\begin{equation}\label{matrixeq}
M_{DDCH}\left(\begin{array}{r}
	b_2 \\
	b_3 \\
	\vdots \\
	b_{14}
\end{array}\right)=v_{DDCH},
\end{equation}
where $M_{DDCH}$ and $v_{DDCH}$ are given by equations (\ref{coeffmatrixddch}) and (\ref{coeffvectors}), respectively.

The matrix equation (\ref{matrixeq}) can be solved analytically to get as many coefficients as needed (not just for 15 coefficients, as given in (\ref{matrixeq})).

Let us redefine,
\begin{equation}
	F_1(u)=A_1 u^2+A_2 u+A_3=\beta u^2+\gamma u+\delta.
\end{equation}
Then the coefficients are obtained as given in the equations (\ref{DDCHcoeffsini}-\ref{DDCHcoeffsfin}).
\begin{figure}[H]
	\centering
	\includegraphics[scale=0.7]{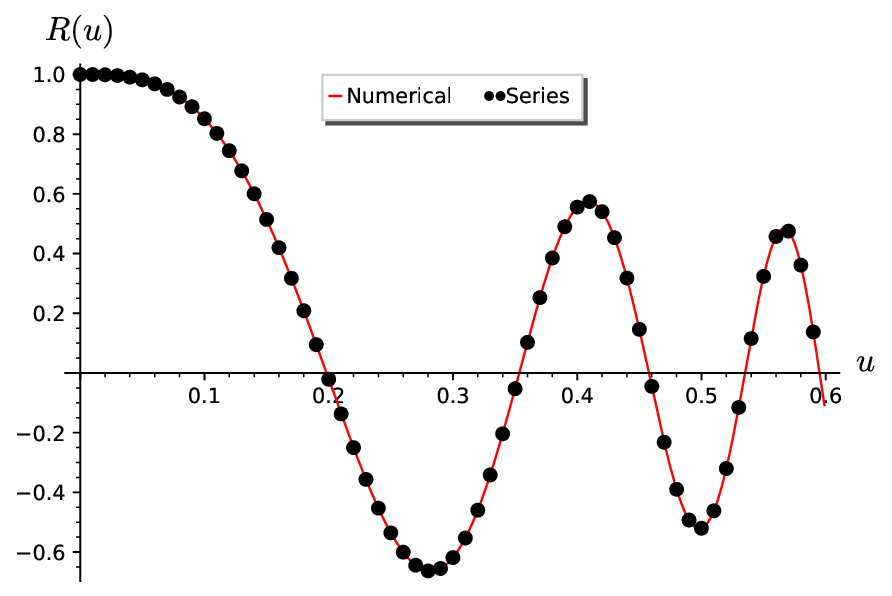} 
	\caption{Numerical solution and the series solution (with 100 terms) of the DDCH equation (\ref{defheund}) with \{$ A_1=\beta=2.06, A_2=\gamma=-895.59, A_3=\delta=1, A_4=-24.3, A_5=-243.0, A_6=-808.677$\}.}
	\label{fig:DDCHcompare}
	\end{figure}
Figure (\ref{fig:DDCHcompare}) shows the agreement of the series solution with the numerical solution. 

All methods for finding the radius of convergence need the calculation of the coefficients ($b_n$) of the series solution as $n \rightarrow \infty$ \cite{gamelin}. The series solution of the DDCH equation is obtained by consecutive Cauchy products of the series. Thus, the computation time for finding the series coefficients grows rapidly for large $n$. While the series solution with $60$ terms can be calculated in a matter of seconds, a solution with $400$ terms can take about two hours even with a powerful computer with an INTEL Core i7 13700KF CPU and 32GB DDR4 memory. 

\begin{figure}[H]
	\begin{subfigure}{.5\textwidth}
		\centering
		% include first image
		\includegraphics[width=1.0\linewidth]{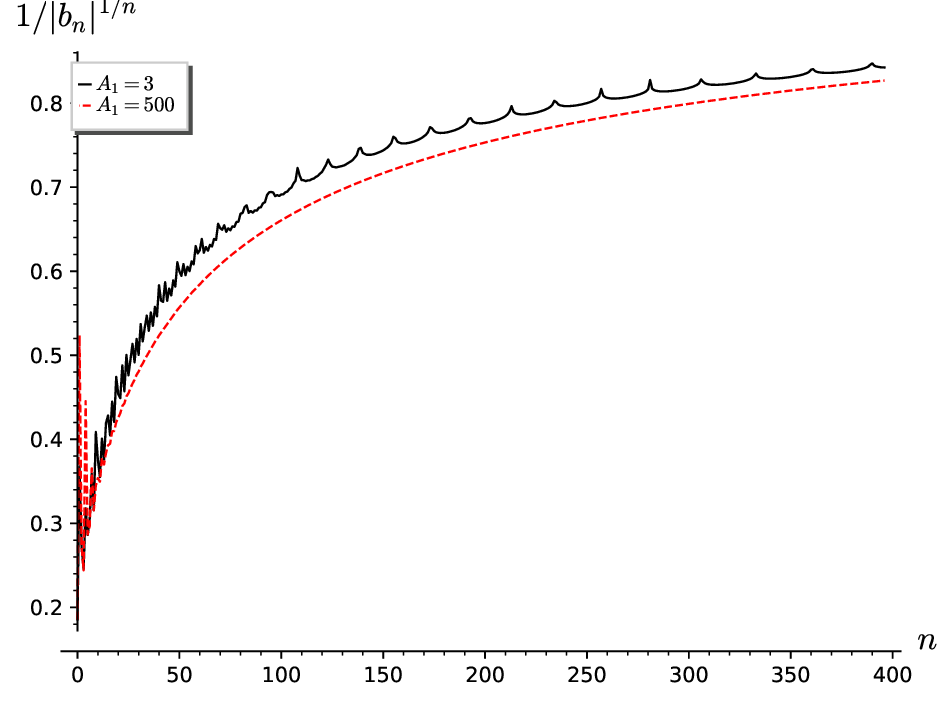}
		\caption{Varying $A_1$}
		\label{fig:DDCH_radofconvA1}
	\end{subfigure}
	\begin{subfigure}{.5\textwidth}
		\centering
		% include first image
		\includegraphics[width=1.0\linewidth]{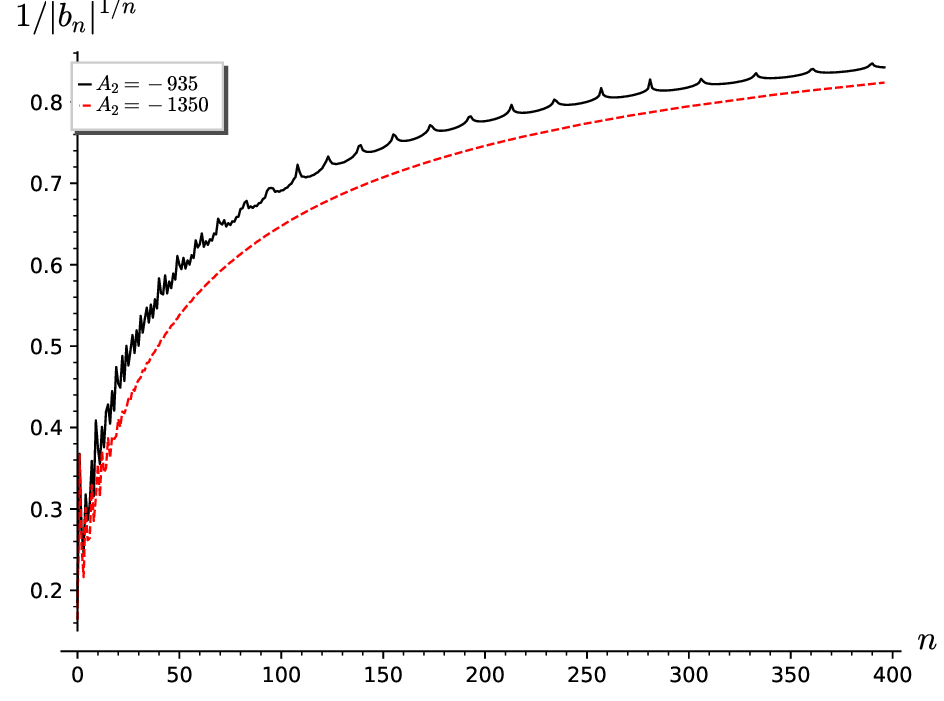}
		\caption{Varying $A_2$}
		\label{fig:DDCH_radofconvA2}
	\end{subfigure}
    \begin{subfigure}{.5\textwidth}
		\centering
		% include second image
		\includegraphics[width=1.0\linewidth]{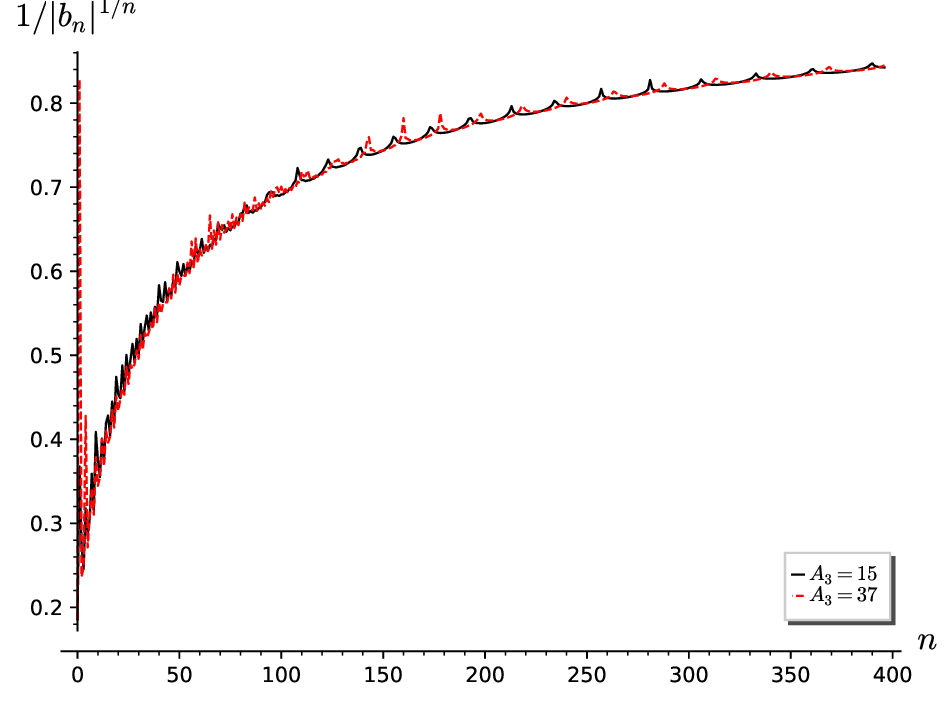}
		\caption{Varying $A_3$}
		\label{fig:DDCH_radofconvA3}
	\end{subfigure}
    \begin{subfigure}{.5\textwidth}
		\centering
		% include second image
		\includegraphics[width=1.0\linewidth]{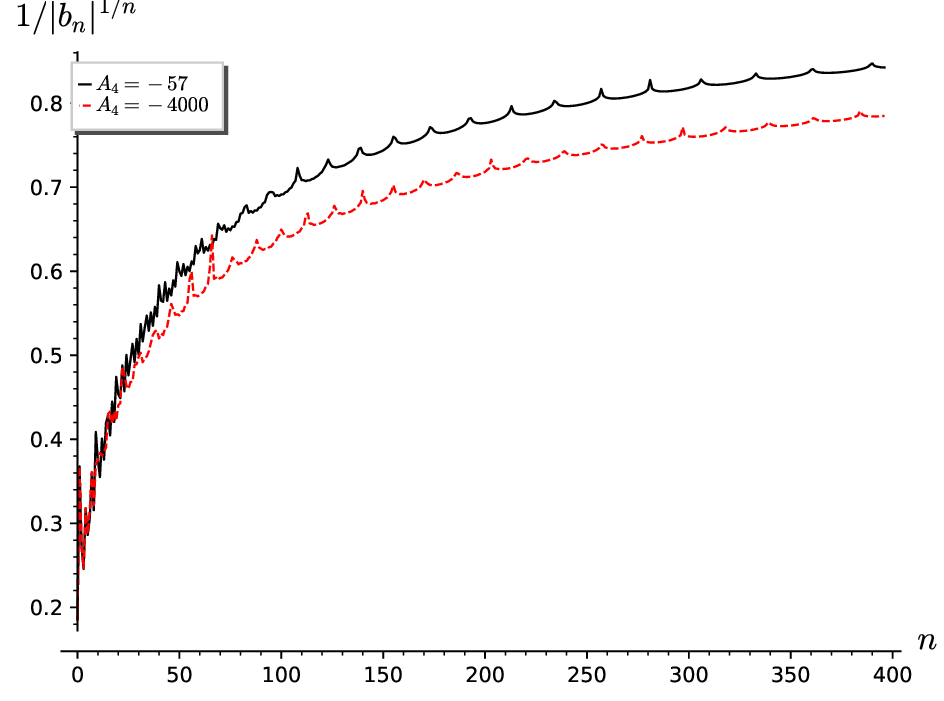}
		\caption{Varying $A_4$}
		\label{fig:DDCH_radofconvA4}
	\end{subfigure}
    \begin{subfigure}{.5\textwidth}
		\centering
		% include second image
		\includegraphics[width=1.0\linewidth]{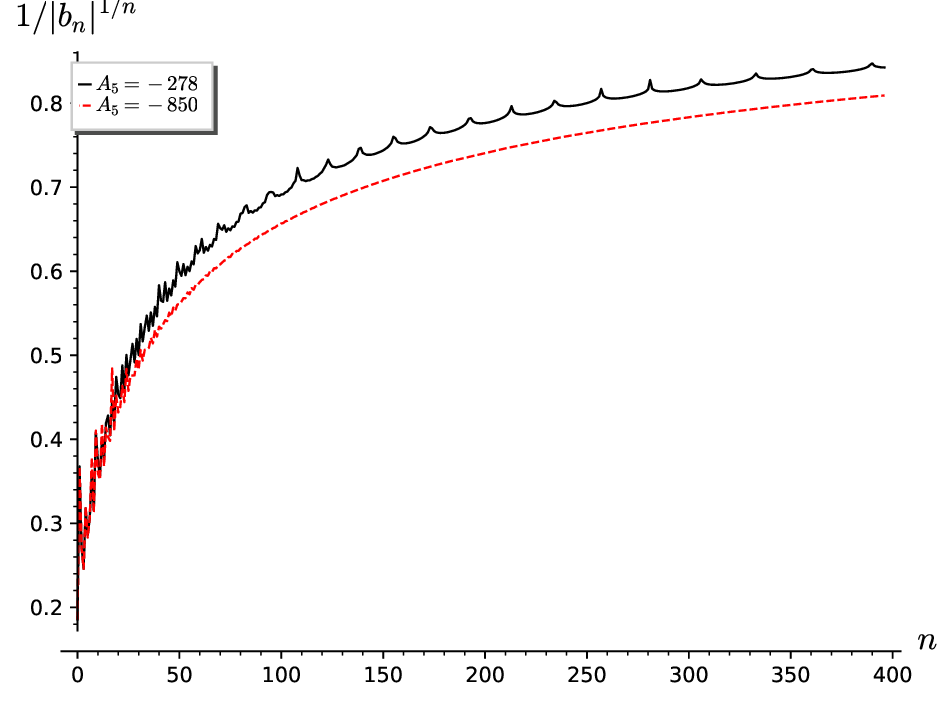}
		\caption{Varying $A_5$}
		\label{fig:DDCH_radofconvA5}
	\end{subfigure}
    \begin{subfigure}{.5\textwidth}
		\centering
		% include second image
		\includegraphics[width=1.0\linewidth]{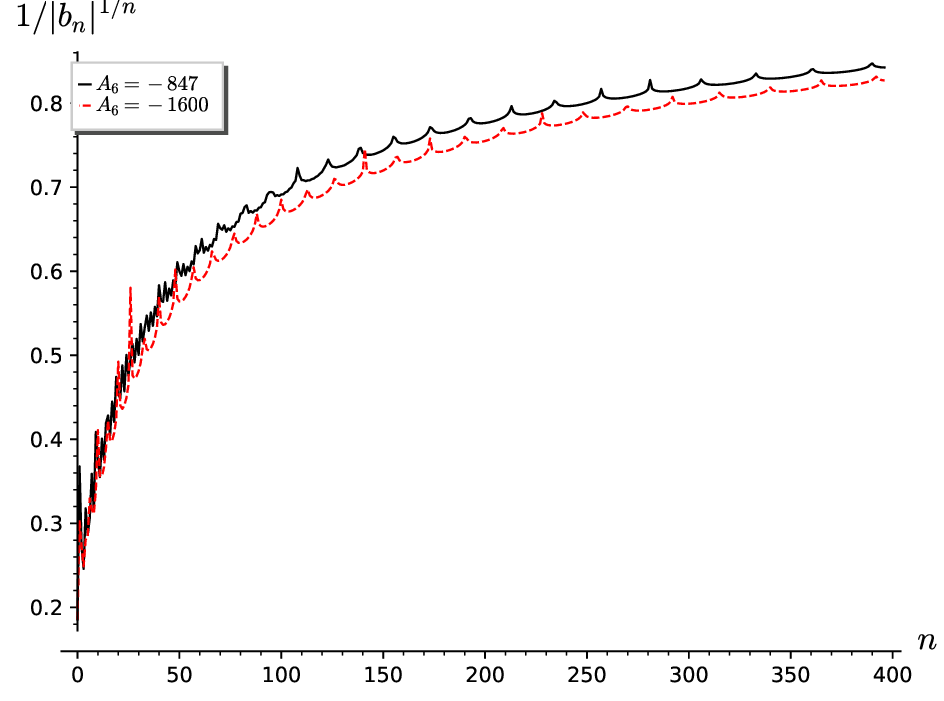}
		\caption{Varying $A_6$}
		\label{fig:DDCH_radofconvA6}
	\end{subfigure}
	\caption{The evolution of $\frac{1}{|b_n|^{1/n}}$ for the radius of convergence. The numerical values of the parameters are \{$ A_1=\beta=3, A_2=\gamma=-935, A_3=\delta=15, A_4=-57, A_5=-278, A_6=-847$\} if not stated otherwise.}
	\label{fig:radconvddch}
\end{figure}

In Figure (\ref{fig:radconvddch}), we analyze the radius of convergence of the DDCH equation with various parameters using the same method we employed in the DCH case. However, we are unable to specify the value of the radius of convergence with only $400$ terms. Yet, the curves seem to grow to a limiting value that is less than the singular point of the DDCH equation located at $u=1$. Therefore we can safely suggest a radius of convergence $|u|<0.7 \sim 0.8$. The observed fluctuations in the curves are most likely due to the numerical calculation of the series coefficients and they start to diminish with increasing $n$. Similar behavior occurs in the DCH case but is less apparent due to the higher $n$ values used in those plots. As can be seen in Appendix A, the DDCH coefficients exhibit a more complicated structure compared to the DCH coefficients.

\begin{figure}[H]
	\begin{subfigure}{.5\textwidth}
		\centering
		% include first image
		\includegraphics[width=1.0\linewidth]{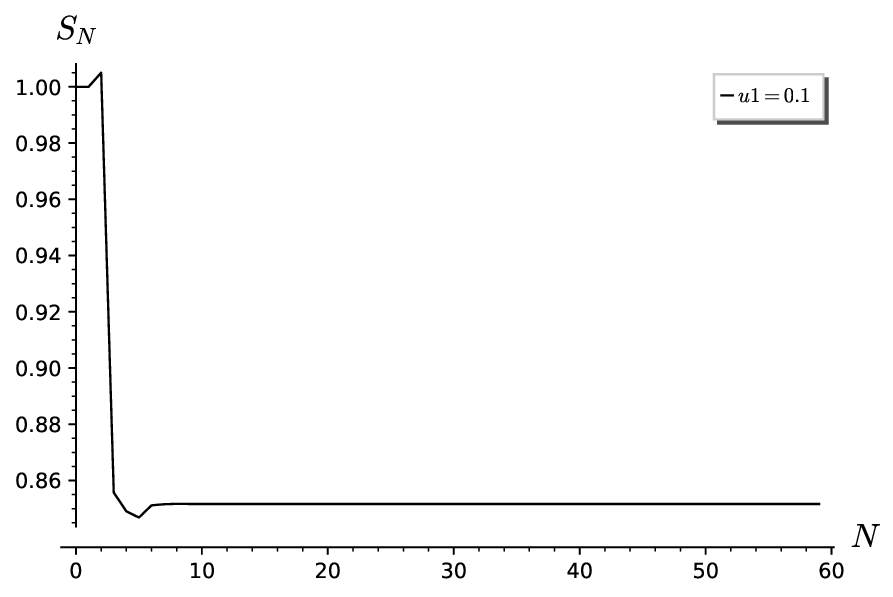}
		\caption{$u=0.1$}
		\label{fig:DDCH_Series_1}
	\end{subfigure}
	\begin{subfigure}{.5\textwidth}
		\centering
		% include second image
		\includegraphics[width=1.0\linewidth]{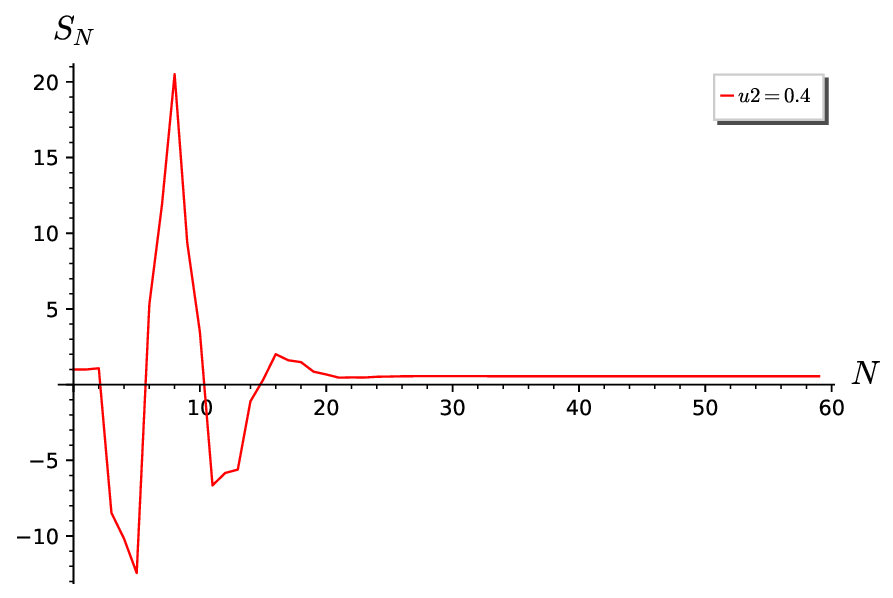}
		\caption{$u=0.4$}
		\label{fig:DDCH_Series_2}
	\end{subfigure}
        \begin{subfigure}{.5\textwidth}
		\centering
		% include second image
		\includegraphics[width=1.0\linewidth]{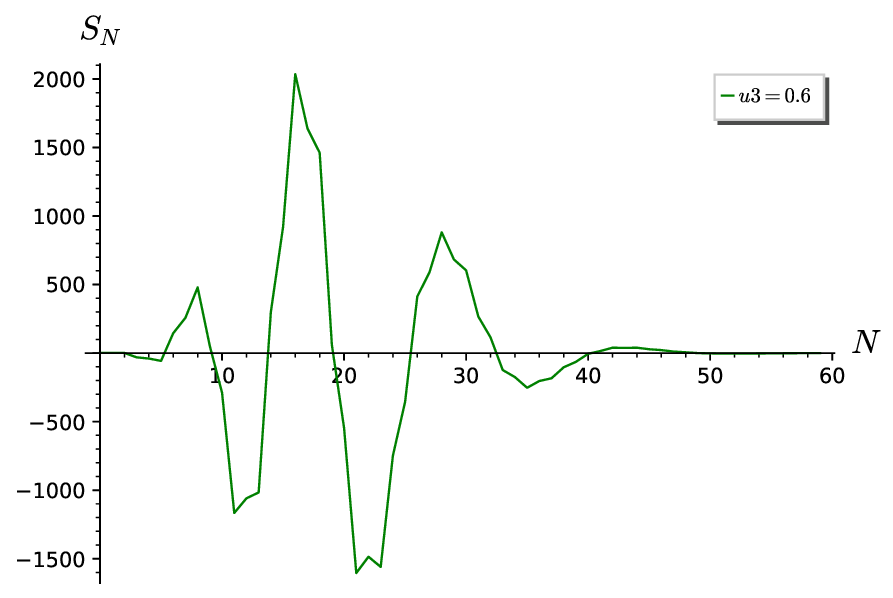}
		\caption{$u=0.6$}
		\label{fig:DDCH_Series_3}
	\end{subfigure}
	\caption{$\bigg(S_N=\sum_{n=0}^{N}b_n u^n$ vs. $N$$\bigg)$. Convergence of the series solution for the DDCH equation (\ref{heund}) with \{$ \beta=2.06, \gamma=-895.59, \delta=1.0 $\}.}
	\label{fig:DCHcompare}
\end{figure}

We can study the convergence of the partial sums of the series as we have done in the previous section. In Figures (\ref{fig:DDCH_Series_1},\ref{fig:DDCH_Series_2},\ref{fig:DDCH_Series_3}), we see that the series solutions asymptotically converge to $S=0.85, 0.55$ and $-0.13$ for $u_1=0.1, u_2=0.4$ and $u_3=0.6$, respectively.  These values are compatible with the numerical result given in Figure (\ref{fig:DDCHcompare}). In all figures, $|S_n - S_m| \rightarrow 0$ as $n, m \rightarrow \infty$ condition is also satisfied.
	
Similar to the DCH case, as $u$ grows larger, more terms are needed in the series for convergence. For $u=0.1$, we need around $10$ terms, for $u=0.4$ we need about $30$ terms and for $u=0.6$ we need around $50$ terms to converge their $S$ values. As mentioned above, the computation time for those number of terms is a few seconds. Thus, our method appears to be effective for equations where the series solution can be studied by convolution applied to the solution ansatz and the coefficients of the equation.
	
In Appendix C, we present the plots of the solutions and their convergence behaviors for all DDCH parameters. We see that the convergence behaviors are independent of the parameters (all converge to $S$ with around 30 terms for $u=0.4$), and series solutions (all with $n=100$ terms) agree with the numerical solutions.
%%%%%%%%%%%%%%%%%%%%%%%%%%%%%%%%%%%%%%%%%%%%%%%%%%%%%%%%%%%%%%%%%%%%%%%%%%%%%%%%%%%%%
\section{Conclusion}
The Klein-Gordon equation in the background of the five-dimensional NGK spacetime yields double-confluent Heun equations for the angular equation and the $c_2=0$ limit of the radial equation. For the full radial equation, we get the ``deformed" double-confluent Heun equation. 

Before delving into the series solution of the new ``deformed" double-confluent Heun equation, we analyzed the formal power series solution of the double-confluent Heun equation and compared it with the numerical solution. We also showed the convergence of the series solutions for several radial values. The equations for the NGK metric forces the parameter $\alpha$ to be zero. We studied both $\alpha=0$ and $\alpha \neq 0$ cases for a complete analysis of the equation.

The ``deformed" double-confluent Heun equation is defined as a double-confluent Heun equation with non-polynomial coefficients. This equation has the same singularity structure as the double-confluent Heun equation although it has no solution in closed form. We employed the Cauchy product for the convolution of the series expansions of the functions present in the coefficients with the power series solution ansatz of the equation. It was impossible to find a recurrence relation between the coefficients, however, we were able to form and solve a matrix equation for the series coefficients analytically and showed that the solution agrees with the numerical solution. We also studied the convergence of the series solution for some radial values.

Physical problems described by Heun-type equations often employ their power series solutions which have convenient recurrence relations. While handling these solutions is simpler than dealing with the equation studied in this article, which involves series convolution and lacks a recurrence relation, we present an equivalent method here with a matrix equation for the coefficients, provided that the convergence of the solution is verified. Thus, our results have the potential to provide a (semi-)analytic investigation of similar problems such as the computation of the connection coefficients needed in the black hole scattering with wave equations having similar non-polynomial structures.
%%%%%%%%%%%%%%%%%%%%%%%%%%%%%%%%%%%%%
\section*{Acknowledgement} 
We would like to thank Profs. Ayşe Hümeyra Bilge, Cihangir Özemir and Yusuf Muhammed Özmen for stimulating discussions. 
%%%%%%%%%%%%%%%%%%%%%%%%%%%%%%%%%%%%%%%%%%%%%%%%%%%%%%%%%%%%%%%%%%%%%%%%%%%%%%%%%%%%%
\begin{landscape}
\section*{Appendix A: Series coefficients for the DCH and DDCH equations} 
For 15 coefficients, the coefficient matrix associated with the DCH equation given in eq. (\ref{heund}) for $\alpha=0$ is
	\tiny
	\begin{equation}\label{coeffmatrixdch}
		M_{DCH}=\left(\begin{array}{rrrrrrrrrrrrr}
			-2 & 0 & 0 & 0 & 0 & 0 & 0 & 0 & 0 & 0 & 0 & 0 & 0 \\
			0 & -6 & 0 & 0 & 0 & 0 & 0 & 0 & 0 & 0 & 0 & 0 & 0 \\
			\delta + 10 & 0 & -12 & 0 & 0 & 0 & 0 & 0 & 0 & 0 & 0 & 0 & 0 \\
			\gamma & \delta + 24 & 0 & -20 & 0 & 0 & 0 & 0 & 0 & 0 & 0 & 0 & 0 \\
			\beta - 14 & \gamma & \delta + 44 & 0 & -30 & 0 & 0 & 0 & 0 & 0 & 0 & 0 & 0 \\
			0 & \beta - 30 & \gamma & \delta + 70 & 0 & -42 & 0 & 0 & 0 & 0 & 0 & 0 & 0 \\
			6 & 0 & \beta - 52 & \gamma & \delta + 102 & 0 & -56 & 0 & 0 & 0 & 0 & 0 & 0 \\
			0 & 12 & 0 & \beta - 80 & \gamma & \delta + 140 & 0 & -72 & 0 & 0 & 0 & 0 & 0 \\
			0 & 0 & 20 & 0 & \beta - 114 & \gamma & \delta + 184 & 0 & -90 & 0 & 0 & 0 & 0 \\
			0 & 0 & 0 & 30 & 0 & \beta - 154 & \gamma & \delta + 234 & 0 & -110 & 0 & 0 & 0 \\
			0 & 0 & 0 & 0 & 42 & 0 & \beta - 200 & \gamma & \delta + 290 & 0 & -132 & 0 & 0 \\
			0 & 0 & 0 & 0 & 0 & 56 & 0 & \beta - 252 & \gamma & \delta + 352 & 0 & -156 & 0 \\
			0 & 0 & 0 & 0 & 0 & 0 & 72 & 0 & \beta - 310 & \gamma & \delta + 420 & 0 & -182
		\end{array}\right),
	\end{equation}
	\normalsize
and the coefficient matrix associated with the DDCH equation given in eq. (\ref{defheund}) is
\tiny
\begin{equation}\label{coeffmatrixddch}
	M_{DDCH}=\left(\begin{array}{rrrrrrrrrrrrr}
		-2 & 0 & 0 & 0 & 0 & 0 & 0 & 0 & 0 & 0 & 0 & 0 & 0 \\
		0 & -6 & 0 & 0 & 0 & 0 & 0 & 0 & 0 & 0 & 0 & 0 & 0 \\
		\delta + 10 & 0 & -12 & 0 & 0 & 0 & 0 & 0 & 0 & 0 & 0 & 0 & 0 \\
		\gamma & \delta + 24 & 0 & -20 & 0 & 0 & 0 & 0 & 0 & 0 & 0 & 0 & 0 \\
		A_{6} + \beta - 14 & \gamma & \delta + 44 & 0 & -30 & 0 & 0 & 0 & 0 & 0 & 0 & 0 & 0 \\
		A_{5} & A_{6} + \beta - 30 & \gamma & \delta + 70 & 0 & -42 & 0 & 0 & 0 & 0 & 0 & 0 & 0 \\
		A_{4} + \frac{1}{3} \, A_{6} + 6 & A_{5} & A_{6} + \beta - 52 & \gamma & \delta + 102 & 0 & -56 & 0 & 0 & 0 & 0 & 0 & 0 \\
		\frac{2}{3} \, A_{5} & A_{4} + \frac{1}{3} \, A_{6} + 12 & A_{5} & A_{6} + \beta - 80 & \gamma & \delta + 140 & 0 & -72 & 0 & 0 & 0 & 0 & 0 \\
		\frac{2}{3} \, A_{4} + \frac{1}{5} \, A_{6} & \frac{2}{3} \, A_{5} & A_{4} + \frac{1}{3} \, A_{6} + 20 & A_{5} & A_{6} + \beta - 114 & \gamma & \delta + 184 & 0 & -90 & 0 & 0 & 0 & 0 \\
		\frac{23}{45} \, A_{5} & \frac{2}{3} \, A_{4} + \frac{1}{5} \, A_{6} & \frac{2}{3} \, A_{5} & A_{4} + \frac{1}{3} \, A_{6} + 30 & A_{5} & A_{6} + \beta - 154 & \gamma & \delta + 234 & 0 & -110 & 0 & 0 & 0 \\
		\frac{28}{45} \, A_{4} + \frac{1}{7} \, A_{6} & \frac{23}{45} \, A_{5} & \frac{2}{3} \, A_{4} + \frac{1}{5} \, A_{6} & \frac{2}{3} \, A_{5} & A_{4} + \frac{1}{3} \, A_{6} + 42 & A_{5} & A_{6} + \beta - 200 & \gamma & \delta + 290 & 0 & -132 & 0 & 0 \\
		\frac{44}{105} \, A_{5} & \frac{28}{45} \, A_{4} + \frac{1}{7} \, A_{6} & \frac{23}{45} \, A_{5} & \frac{2}{3} \, A_{4} + \frac{1}{5} \, A_{6} & \frac{2}{3} \, A_{5} & A_{4} + \frac{1}{3} \, A_{6} + 56 & A_{5} & A_{6} + \beta - 252 & \gamma & \delta + 352 & 0 & -156 & 0 \\
		\frac{494}{945} \, A_{4} + \frac{1}{9} \, A_{6} & \frac{44}{105} \, A_{5} & \frac{28}{45} \, A_{4} + \frac{1}{7} \, A_{6} & \frac{23}{45} \, A_{5} & \frac{2}{3} \, A_{4} + \frac{1}{5} \, A_{6} & \frac{2}{3} \, A_{5} & A_{4} + \frac{1}{3} \, A_{6} + 72 & A_{5} & A_{6} + \beta - 310 & \gamma & \delta + 420 & 0 & -182
	\end{array}\right),
\end{equation}
\normalsize
and the vectors,
\begin{align}\label{coeffvectors}
	v_{DCH}=\left(\begin{array}{r}
		-\delta \\
		-\gamma \\
		-\beta \\
		0 \\
		0 \\
		0 \\
		0 \\
		0 \\
		0 \\
		0 \\
		0 \\
		0 \\
		0
	\end{array}\right), & \quad\quad\quad\quad
	v_{DDCH}=\left(\begin{array}{r}
		-\delta \\
		-\gamma \\
		-A_{6} - \beta \\
		-A_{5} \\
		-A_{4} - \frac{1}{3} \, A_{6} \\
		-\frac{2}{3} \, A_{5} \\
		-\frac{2}{3} \, A_{4} - \frac{1}{5} \, A_{6} \\
		-\frac{23}{45} \, A_{5} \\
		-\frac{28}{45} \, A_{4} - \frac{1}{7} \, A_{6} \\
		-\frac{44}{105} \, A_{5} \\
		-\frac{494}{945} \, A_{4} - \frac{1}{9} \, A_{6} \\
		-\frac{563}{1575} \, A_{5} \\
		-\frac{257}{525} \, A_{4} - \frac{1}{11} \, A_{6}
	\end{array}\right)
\end{align}
The matrices and vectors can be generalized easily for more terms using the series solution obtained in the text.

The coefficients for the DCH equation with $\alpha=0$ are found as
\tiny
\begin{align}\label{DCHcoeffsini}
b_0&=1, b_1=0, b_2=\frac{1}{2} \, \delta, b_3=\frac{1}{6} \, \gamma, b_4= \frac{1}{24} \, \delta^{2} + \frac{1}{12} \, \beta + \frac{5}{12} \, \delta, b_5= \frac{1}{30} \, {\left(\delta + 6\right)} \gamma\\
b_6&= \frac{1}{720} \, \delta^{3} + \frac{1}{360} \, {\left(7 \, \beta + 136\right)} \delta + \frac{3}{40} \, \delta^{2} + \frac{1}{180} \, \gamma^{2} + \frac{11}{90} \, \beta \\
b_7&= \frac{1}{1680} \, {\left(3 \, \delta^{2} + 10 \, \beta + 118 \, \delta + 360\right)} \gamma\\
b_8&= \frac{1}{40320} \, \delta^{4} + \frac{1}{10080} \, {\left(11 \, \beta + 1055\right)} \delta^{2} + \frac{13}{3360} \, \delta^{3} + \frac{1}{10080} \, {\left(7 \, \delta + 138\right)} \gamma^{2} + \frac{1}{672} \, \beta^{2} + \frac{1}{5040} \, {\left(227 \, \beta + 1788\right)} \delta + \frac{61}{420} \, \beta\\
b_9&= \frac{1}{12960} \, \gamma^{3} + \frac{1}{45360} \, {\left(2 \, \delta^{3} + {\left(37 \, \beta + 4888\right)} \delta + 249 \, \delta^{2} + 728 \, \beta + 10080\right)} \gamma\\
b_{10}&= \frac{1}{3628800} \, \delta^{5} + \frac{1}{181440} \, {\left(5 \, \beta + 1327\right)} \delta^{3} + \frac{17}{181440} \, \delta^{4} + \frac{1}{453600} \, {\left(1617 \, \beta + 59956\right)} \delta^{2} + \frac{1}{907200} \, {\left(25 \, \delta^{2} + 116 \, \beta + 2134 \, \delta + 21168\right)} \gamma^{2} + \frac{499}{113400} \, \beta^{2} + \frac{1}{907200} \, {\left(211 \, \beta^{2} + 66464 \, \beta + 307872\right)} \delta + \frac{253}{1575} \, \beta\\ 
b_{11}&= \frac{1}{997920} \, {\left(7 \, \delta + 288\right)} \gamma^{3} + \frac{1}{7983360} \, {\left(5 \, \delta^{4} + 4 \, {\left(67 \, \beta + 22171\right)} \delta^{2} + 1428 \, \delta^{3} + 540 \, \beta^{2} + 8 \, {\left(2923 \, \beta + 144936\right)} \delta + 232128 \, \beta + 1814400\right)} \gamma\\
b_{12}&= \frac{1}{479001600} \, \delta^{6} + \frac{1}{47900160} \, {\left(19 \, \beta + 10714\right)} \delta^{4} + \frac{1}{760320} \, \delta^{5} + \frac{1}{59875200} \, {\left(6997 \, \beta + 697491\right)} \delta^{3} + \frac{1}{1710720} \, \gamma^{4} + \frac{1}{88704} \, \beta^{3} + \frac{1}{119750400} \, {\left(1201 \, \beta^{2} + 901274 \, \beta + 18950032\right)} \delta^{2} \nonumber \\
&+ \frac{1}{119750400} \, {\left(65 \, \delta^{3} + 2 \, {\left(743 \, \beta + 305894\right)} \delta + 14364 \, \delta^{2} + 60620 \, \beta + 4068000\right)} \gamma^{2} + \frac{12743}{1496880} \, \beta^{2} + \frac{1}{59875200} \, {\left(53021 \, \beta^{2} + 6155504 \, \beta + 19654560\right)} \delta + \frac{7141}{41580} \, \beta\\
b_{13}&= \frac{1}{172972800} \, \delta^{5} \gamma + \frac{1}{1482624} \, \beta \delta^{3} \gamma + \frac{41}{12972960} \, \delta^{4} \gamma + \frac{23}{103783680} \, \delta^{2} \gamma^{3} \nonumber \\
&+ \frac{29}{4054050} \, \beta^{2} \delta \gamma + \frac{1237}{8108100} \, \beta \delta^{2} \gamma + \frac{487}{1081080} \, \delta^{3} \gamma + \frac{31}{23587200} \, \beta \gamma^{3} + \frac{2833}{86486400} \, \delta \gamma^{3} + \frac{239}{842400} \, \beta^{2} \gamma + \frac{1329}{200200} \, \beta \delta \gamma + \frac{109783}{5896800} \, \delta^{2} \gamma + \frac{443}{655200} \, \gamma^{3} + \frac{259}{5850} \, \beta \gamma + \frac{41101}{225225} \, \delta \gamma + \frac{3}{13} \, \gamma\\
b_{14}&= \frac{1}{87178291200} \, \delta^{7} + \frac{23}{6227020800} \, \beta \delta^{5} + \frac{1}{83026944} \, \delta^{6} + \frac{1}{155675520} \, \delta^{4} \gamma^{2} + \frac{643}{3113510400} \, \beta^{2} \delta^{3} + \frac{1613}{778377600} \, \beta \delta^{4} + \frac{2363}{622702080} \, \delta^{5} + \frac{629}{1556755200} \, \beta \delta^{2} \gamma^{2} + \frac{751}{259459200} \, \delta^{3} \gamma^{2} \nonumber \\
&+ \frac{1}{23950080} \, \delta \gamma^{4} + \frac{4867}{3632428800} \, \beta^{3} \delta + \frac{20747}{435891456} \, \beta^{2} \delta^{2} + \frac{3315787}{10897286400} \, \beta \delta^{3} + \frac{74447}{172972800} \, \delta^{4} + \frac{1951}{1816214400} \, \beta^{2} \gamma^{2} + \frac{329297}{5448643200} \, \beta \delta \gamma^{2} + \frac{868991}{2724321600} \, \delta^{2} \gamma^{2} \nonumber \\
&+ \frac{41}{13970880} \, \gamma^{4} + \frac{45581}{908107200} \, \beta^{3} + \frac{2856013}{1362160800} \, \beta^{2} \delta + \frac{35468519}{2724321600} \, \beta \delta^{2} + \frac{11459291}{681080400} \, \delta^{3} + \frac{17039}{13759200} \, \beta \gamma^{2} + \frac{68429}{7567560} \, \delta \gamma^{2} \nonumber \\
&+ \frac{1546243}{113513400} \, \beta^{2} + \frac{15106537}{113513400} \, \beta \delta + \frac{2080171}{11351340} \, \delta^{2} + \frac{28577}{630630} \, \gamma^{2} + \frac{113623}{630630} \, \beta + \frac{100846}{315315} \, \delta \label{DCHcoeffsfin} .
\end{align}
\normalsize
and the coefficients for the DDCH equation are
\tiny
\begingroup
\allowdisplaybreaks
\begin{align}\label{DDCHcoeffsini}
b_0&= 1, b_1= 0, b_2= \frac{1}{2} \, \delta, b_3= \frac{1}{6} \, \gamma, b_4= \frac{1}{24} \, \delta^{2} + \frac{1}{12} \, A_{6} + \frac{1}{12} \, \beta + \frac{5}{12} \, \delta, b_5= \frac{1}{30} \, {\left(\delta + 6\right)} \gamma + \frac{1}{20} \, A_{5}\\
b_6&= \frac{1}{720} \, \delta^{3} + \frac{1}{360} \, {\left(7 \, A_{6} + 7 \, \beta + 136\right)} \delta + \frac{3}{40} \, \delta^{2} + \frac{1}{180} \, \gamma^{2} + \frac{1}{30} \, A_{4} + \frac{2}{15} \, A_{6} + \frac{11}{90} \, \beta\\
b_7&=\frac{11}{840} \, A_{5} \delta + \frac{1}{1680} \, {\left(3 \, \delta^{2} + 10 \, A_{6} + 10 \, \beta + 118 \, \delta + 360\right)} \gamma + \frac{25}{252} \, A_{5}\\
b_8&=\frac{1}{40320} \, \delta^{4} + \frac{1}{10080} \, {\left(11 \, A_{6} + 11 \, \beta + 1055\right)} \delta^{2} + \frac{13}{3360} \, \delta^{3} + \frac{1}{10080} \, {\left(7 \, \delta + 138\right)} \gamma^{2} + \frac{1}{672} \, A_{6}^{2} + \frac{1}{1680} \, {\left(5 \, A_{6} + 244\right)} \beta + \frac{1}{672} \, \beta^{2} + \frac{1}{5040} \, {\left(48 \, A_{4} + 243 \, A_{6} + 227 \, \beta + 1788\right)} \delta + \frac{13}{3360} \, A_{5} \gamma + \frac{61}{840} \, A_{4} + \frac{71}{420} \, A_{6} \\
b_9&=\frac{23}{30240} \, A_{5} \delta^{2} + \frac{1}{12960} \, \gamma^{3} + \frac{1}{540} \, A_{5} A_{6} + \frac{1}{540} \, A_{5} \beta + \frac{169}{4536} \, A_{5} \delta + \frac{1}{45360} \, {\left(2 \, \delta^{3} + {\left(37 \, A_{6} + 37 \, \beta + 4888\right)} \delta + 249 \, \delta^{2} + 126 \, A_{4} + 770 \, A_{6} + 728 \, \beta + 10080\right)} \gamma + \frac{13}{90} \, A_{5}\\
b_{10}&= \frac{1}{3628800} \, \delta^{5} + \frac{1}{181440} \, {\left(5 \, A_{6} + 5 \, \beta + 1327\right)} \delta^{3} + \frac{17}{181440} \, \delta^{4} + \frac{1}{453600} \, {\left(258 \, A_{4} + 1703 \, A_{6} + 1617 \, \beta + 59956\right)} \delta^{2} \nonumber \\
&+ \frac{1}{907200} \, {\left(25 \, \delta^{2} + 116 \, A_{6} + 116 \, \beta + 2134 \, \delta + 21168\right)} \gamma^{2} + \frac{1}{1800} \, A_{5}^{2} + \frac{1}{37800} \, {\left(49 \, A_{4} + 7440\right)} A_{6} + \frac{137}{28350} \, A_{6}^{2} + \frac{1}{37800} \, {\left(49 \, A_{4} + 349 \, A_{6} + 6072\right)} \beta + \frac{499}{113400} \, \beta^{2} \nonumber \\
&+ \frac{1}{907200} \, {\left(211 \, A_{6}^{2} + 2 \, {\left(211 \, A_{6} + 33232\right)} \beta + 211 \, \beta^{2} + 25956 \, A_{4} + 75000 \, A_{6} + 307872\right)} \delta + \frac{1}{907200} \, {\left(507 \, A_{5} \delta + 11312 \, A_{5}\right)} \gamma + \frac{1604}{14175} \, A_{4}\\ 
b_{11}&= \frac{13}{665280} \, A_{5} \delta^{3} + \frac{1}{997920} \, {\left(7 \, \delta + 288\right)} \gamma^{3} + \frac{577}{199584} \, A_{5} \delta^{2} + \frac{19}{221760} \, A_{5} \gamma^{2} + \frac{1}{138600} \, {\left(105 \, A_{4} + 25756\right)} A_{5} + \frac{31}{4620} \, A_{5} A_{6} + \frac{179}{27720} \, A_{5} \beta + \frac{1}{207900} \, {\left(65 \, A_{5} A_{6} + 65 \, A_{5} \beta + 14661 \, A_{5}\right)} \delta \nonumber \\
&+ \frac{1}{39916800} \, {\left(25 \, \delta^{4} + 20 \, {\left(67 \, A_{6} + 67 \, \beta + 22171\right)} \delta^{2} + 7140 \, \delta^{3} + 2700 \, A_{6}^{2} + 360 \, {\left(15 \, A_{6} + 3224\right)} \beta + 2700 \, \beta^{2} + 40 \, {\left(414 \, A_{4} + 3061 \, A_{6} + 2923 \, \beta + 144936\right)} \delta + 375120 \, A_{4} + 1284192 \, A_{6} + 9072000\right)} \gamma\\
b_{12}&= \frac{1}{479001600} \, \delta^{6} + \frac{1}{47900160} \, {\left(19 \, A_{6} + 19 \, \beta + 10714\right)} \delta^{4} + \frac{1}{760320} \, \delta^{5} + \frac{1}{59875200} \, {\left(888 \, A_{4} + 7293 \, A_{6} + 6997 \, \beta + 697491\right)} \delta^{3} \nonumber \\
&+ \frac{1}{1710720} \, \gamma^{4} + \frac{1}{88704} \, A_{6}^{3} + \frac{1}{11975040} \, {\left(405 \, A_{6} + 101944\right)} \beta^{2} + \frac{1}{88704} \, \beta^{3} + \frac{1}{119750400} \, {\left(1201 \, A_{6}^{2} + 2 \, {\left(1201 \, A_{6} + 450637\right)} \beta + 1201 \, \beta^{2} + 268836 \, A_{4} + 989930 \, A_{6} + 18950032\right)} \delta^{2} \nonumber \\
&+ \frac{1}{119750400} \, {\left(65 \, \delta^{3} + 2 \, {\left(743 \, A_{6} + 743 \, \beta + 305894\right)} \delta + 14364 \, \delta^{2} + 7560 \, A_{4} + 63140 \, A_{6} + 60620 \, \beta + 4068000\right)} \gamma^{2} \nonumber \\
&+ \frac{1}{3960} \, A_{4}^{2} + \frac{37}{16632} \, A_{5}^{2} + \frac{1}{997920} \, {\left(4903 \, A_{4} + 219096\right)} A_{6} + \frac{2161}{213840} \, A_{6}^{2} + \frac{1}{3991680} \, {\left(135 \, A_{6}^{2} + 18940 \, A_{4} + 74208 \, A_{6} + 685536\right)} \beta \nonumber \\
&+ \frac{1}{59875200} \, {\left(6192 \, A_{5}^{2} + 96 \, {\left(143 \, A_{4} + 75510\right)} A_{6} + 57597 \, A_{6}^{2} + 2 \, {\left(6864 \, A_{4} + 55309 \, A_{6} + 3077752\right)} \beta + 53021 \, \beta^{2} + 3389428 \, A_{4} + 19654560\right)} \delta \nonumber \\
&+ \frac{1}{119750400} \, {\left(2817 \, A_{5} \delta^{2} + 10590 \, A_{5} A_{6} + 10590 \, A_{5} \beta + 276022 \, A_{5} \delta + 3102160 \, A_{5}\right)} \gamma + \frac{14327}{93555} \, A_{4}\\
b_{13}&=\frac{1}{172972800} \, \delta^{5} \gamma + \frac{59}{207567360} \, A_{5} \delta^{4} + \frac{1}{1482624} \, A_{6} \delta^{3} \gamma + \frac{1}{1482624} \, \beta \delta^{3} \gamma + \frac{41}{12972960} \, \delta^{4} \gamma + \frac{23}{103783680} \, \delta^{2} \gamma^{3} + \frac{1}{72072} \, A_{5} A_{6} \delta^{2} + \frac{1}{72072} \, A_{5} \beta \delta^{2} \nonumber \\
&+ \frac{559}{5987520} \, A_{5} \delta^{3} + \frac{29}{4054050} \, A_{6}^{2} \delta \gamma + \frac{29}{2027025} \, A_{6} \beta \delta \gamma + \frac{29}{4054050} \, \beta^{2} \delta \gamma + \frac{329}{18532800} \, A_{4} \delta^{2} \gamma + \frac{61679}{389188800} \, A_{6} \delta^{2} \gamma + \frac{1237}{8108100} \, \beta \delta^{2} \gamma \nonumber \\
&+ \frac{487}{1081080} \, \delta^{3} \gamma + \frac{2227}{259459200} \, A_{5} \delta \gamma^{2} + \frac{31}{23587200} \, A_{6} \gamma^{3} + \frac{31}{23587200} \, \beta \gamma^{3} + \frac{2833}{86486400} \, \delta \gamma^{3} + \frac{101}{4717440} \, A_{5} A_{6}^{2} + \frac{101}{2358720} \, A_{5} A_{6} \beta + \frac{101}{4717440} \, A_{5} \beta^{2} \nonumber \\
&+ \frac{3}{20020} \, A_{4} A_{5} \delta + \frac{109553}{77837760} \, A_{5} A_{6} \delta + \frac{3019}{2223936} \, A_{5} \beta \delta + \frac{5351089}{778377600} \, A_{5} \delta^{2} + \frac{223}{7862400} \, A_{5}^{2} \gamma + \frac{379}{5896800} \, A_{4} A_{6} \gamma + \frac{2699}{8845200} \, A_{6}^{2} \gamma + \frac{379}{5896800} \, A_{4} \beta \gamma \nonumber \\
&+ \frac{10417}{17690400} \, A_{6} \beta \gamma + \frac{239}{842400} \, \beta^{2} \gamma + \frac{38323}{21621600} \, A_{4} \delta \gamma + \frac{1405627}{194594400} \, A_{6} \delta \gamma + \frac{1329}{200200} \, \beta \delta \gamma + \frac{109783}{5896800} \, \delta^{2} \gamma + \frac{13613}{35380800} \, A_{5} \gamma^{2} + \frac{443}{655200} \, \gamma^{3} + \frac{83}{26208} \, A_{4} A_{5} + \frac{11261}{737100} \, A_{5} A_{6} \nonumber \\
&+ \frac{1049}{73710} \, A_{5} \beta + \frac{138991}{1247400} \, A_{5} \delta + \frac{44969}{2211300} \, A_{4} \gamma + \frac{2777}{54600} \, A_{6} \gamma + \frac{259}{5850} \, \beta \gamma + \frac{41101}{225225} \, \delta \gamma + \frac{55007}{245700} \, A_{5} + \frac{3}{13} \, \gamma\\
b_{14}&=\frac{1}{87178291200} \, \delta^{7} + \frac{23}{6227020800} \, A_{6} \delta^{5} + \frac{23}{6227020800} \, \beta \delta^{5} + \frac{1}{83026944} \, \delta^{6} + \frac{1}{155675520} \, \delta^{4} \gamma^{2} + \frac{643}{3113510400} \, A_{6}^{2} \delta^{3} + \frac{643}{1556755200} \, A_{6} \beta \delta^{3} + \frac{643}{3113510400} \, \beta^{2} \delta^{3} + \frac{113}{518918400} \, A_{4} \delta^{4} + \frac{53}{24710400} \, A_{6} \delta^{4} + \frac{1613}{778377600} \, \beta \delta^{4} \nonumber \\
&+ \frac{2363}{622702080} \, \delta^{5} + \frac{71}{148262400} \, A_{5} \delta^{3} \gamma + \frac{629}{1556755200} \, A_{6} \delta^{2} \gamma^{2} + \frac{629}{1556755200} \, \beta \delta^{2} \gamma^{2} + \frac{751}{259459200} \, \delta^{3} \gamma^{2} + \frac{1}{23950080} \, \delta \gamma^{4} \nonumber \\
&+ \frac{4867}{3632428800} \, A_{6}^{3} \delta + \frac{4867}{1210809600} \, A_{6}^{2} \beta \delta + \frac{4867}{1210809600} \, A_{6} \beta^{2} \delta + \frac{4867}{3632428800} \, \beta^{3} \delta + \frac{479}{100900800} \, A_{5}^{2} \delta^{2} + \frac{857}{82555200} \, A_{4} A_{6} \delta^{2} + \frac{185461}{3632428800} \, A_{6}^{2} \delta^{2} \nonumber \\
&+ \frac{857}{82555200} \, A_{4} \beta \delta^{2} + \frac{537529}{5448643200} \, A_{6} \beta \delta^{2} + \frac{20747}{435891456} \, \beta^{2} \delta^{2} + \frac{132413}{1816214400} \, A_{4} \delta^{3} + \frac{3578287}{10897286400} \, A_{6} \delta^{3} + \frac{3315787}{10897286400} \, \beta \delta^{3} \nonumber \\
&+ \frac{74447}{172972800} \, \delta^{4} + \frac{11813}{1210809600} \, A_{5} A_{6} \delta \gamma + \frac{11813}{1210809600} \, A_{5} \beta \delta \gamma + \frac{1302641}{10897286400} \, A_{5} \delta^{2} \gamma + \frac{1951}{1816214400} \, A_{6}^{2} \gamma^{2} + \frac{1951}{908107200} \, A_{6} \beta \gamma^{2} + \frac{1951}{1816214400} \, \beta^{2} \gamma^{2} + \frac{1}{155232} \, A_{4} \delta \gamma^{2} + \frac{340997}{5448643200} \, A_{6} \delta \gamma^{2} \nonumber \\
&+ \frac{329297}{5448643200} \, \beta \delta \gamma^{2} + \frac{868991}{2724321600} \, \delta^{2} \gamma^{2} + \frac{5}{5588352} \, A_{5} \gamma^{3} + \frac{41}{13970880} \, \gamma^{4} + \frac{1}{75600} \, A_{5}^{2} A_{6} + \frac{421}{27518400} \, A_{4} A_{6}^{2} + \frac{12553}{227026800} \, A_{6}^{3} + \frac{1}{75600} \, A_{5}^{2} \beta + \frac{421}{13759200} \, A_{4} A_{6} \beta + \frac{29201}{181621440} \, A_{6}^{2} \beta + \frac{421}{27518400} \, A_{4} \beta^{2} \nonumber \\
&+ \frac{70687}{454053600} \, A_{6} \beta^{2} + \frac{45581}{908107200} \, \beta^{3} + \frac{271}{5045040} \, A_{4}^{2} \delta + \frac{1429}{2837835} \, A_{5}^{2} \delta + \frac{968843}{908107200} \, A_{4} A_{6} \delta + \frac{1663993}{681080400} \, A_{6}^{2} \delta + \frac{936323}{908107200} \, A_{4} \beta \delta + \frac{294089}{64864800} \, A_{6} \beta \delta + \frac{2856013}{1362160800} \, \beta^{2} \delta + \frac{7512551}{1362160800} \, A_{4} \delta^{2} + \frac{1019}{68796} \, A_{6} \delta^{2} + \frac{35468519}{2724321600} \, \beta \delta^{2} \nonumber \\
&+ \frac{11459291}{681080400} \, \delta^{3} + \frac{821}{20180160} \, A_{4} A_{5} \gamma + \frac{32663}{75675600} \, A_{5} A_{6} \gamma + \frac{126547}{302702400} \, A_{5} \beta \gamma + \frac{3977549}{681080400} \, A_{5} \delta \gamma + \frac{31}{105840} \, A_{4} \gamma^{2} + \frac{1943}{1455300} \, A_{6} \gamma^{2} + \frac{17039}{13759200} \, \beta \gamma^{2} + \frac{68429}{7567560} \, \delta \gamma^{2} + \frac{5569}{5045040} \, A_{4}^{2} \nonumber \\
&+ \frac{103769}{18918900} \, A_{5}^{2} + \frac{1319449}{113513400} \, A_{4} A_{6} + \frac{245209}{14189175} \, A_{6}^{2} + \frac{353321}{32432400} \, A_{4} \beta + \frac{1164743}{37837800} \, A_{6} \beta + \frac{1546243}{113513400} \, \beta^{2} + \frac{157847}{1702701} \, A_{4} \delta + \frac{3072701}{18918900} \, A_{6} \delta \nonumber \\
&+ \frac{15106537}{113513400} \, \beta \delta + \frac{2080171}{11351340} \, \delta^{2} + \frac{624971}{14189175} \, A_{5} \gamma + \frac{28577}{630630} \, \gamma^{2} + \frac{2725427}{14189175} \, A_{4} + \frac{10756}{45045} \, A_{6} + \frac{113623}{630630} \, \beta + \frac{100846}{315315} \, \delta. \label{DDCHcoeffsfin}
\end{align}
\endgroup
\normalsize

\end{landscape}
%%%%%%%%%%%%%%%%%%%%%%%%%%%%%%%%%%%%%%%%%%%%%%%%%%%%%%%%%%%%%%%%%%%%%%%%%%%%%%%%%%%%%
\section*{Appendix B: Series Solution Analysis of the DCH Equation for Various Parameters}
\begin{figure}[H]
	\begin{subfigure}{.5\textwidth}
		\centering
		% include first image
		\includegraphics[width=1.0\linewidth]{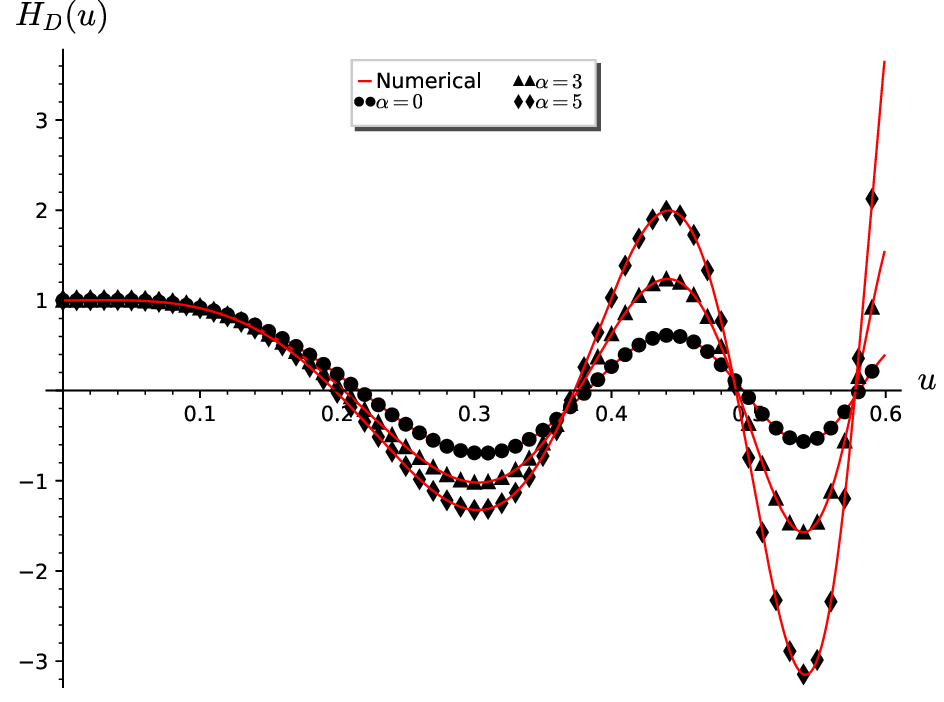}
		\caption{$H_D$ for varying $\alpha$}
		\label{fig:DCH_alpha}
	\end{subfigure}
	\begin{subfigure}{.5\textwidth}
		\centering
		% include second image
		\includegraphics[width=1.0\linewidth]{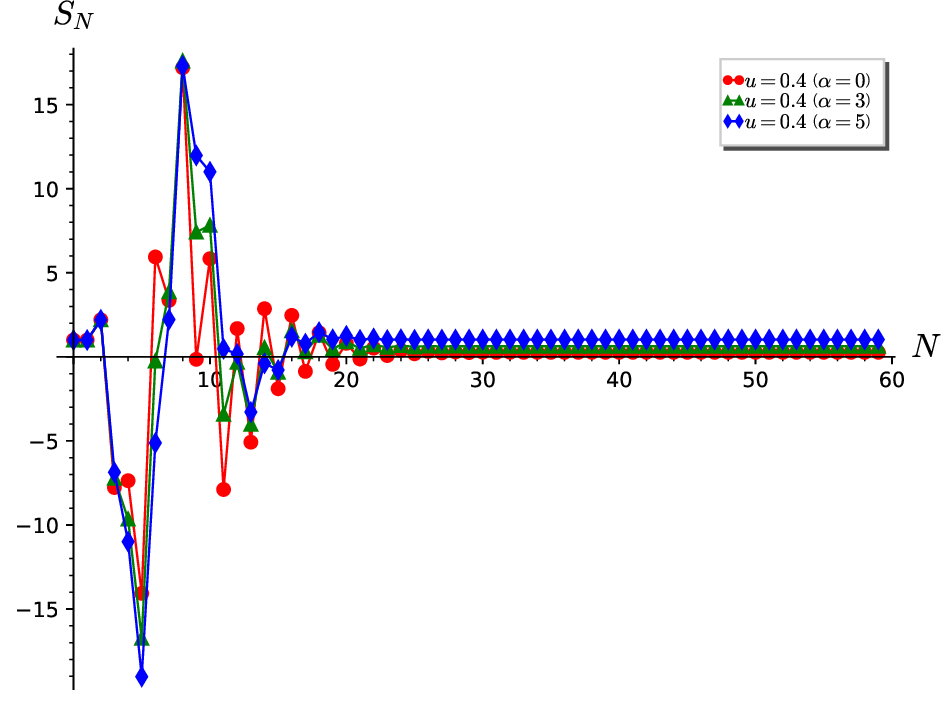}
		\caption{Convergence for $u=0.4$}
		\label{fig:DCH_conv_alpha}
	\end{subfigure}
\caption{ Analysis of the DCH series solution for varying $\alpha$ with \{$ \beta=3, \gamma=-935, \delta=15 $\}.} 
\end{figure}
As seen from the Figure (\ref{fig:DCH_conv_alpha}), the series solutions converge to $S=0.27,0.63$ and $1.03$ for $\alpha=0, \alpha=3$ and $\alpha=5$ at $u=0.4$, respectively.
\begin{figure}[H]
	\begin{subfigure}{.5\textwidth}
		\centering
		% include first image
		\includegraphics[width=1.0\linewidth]{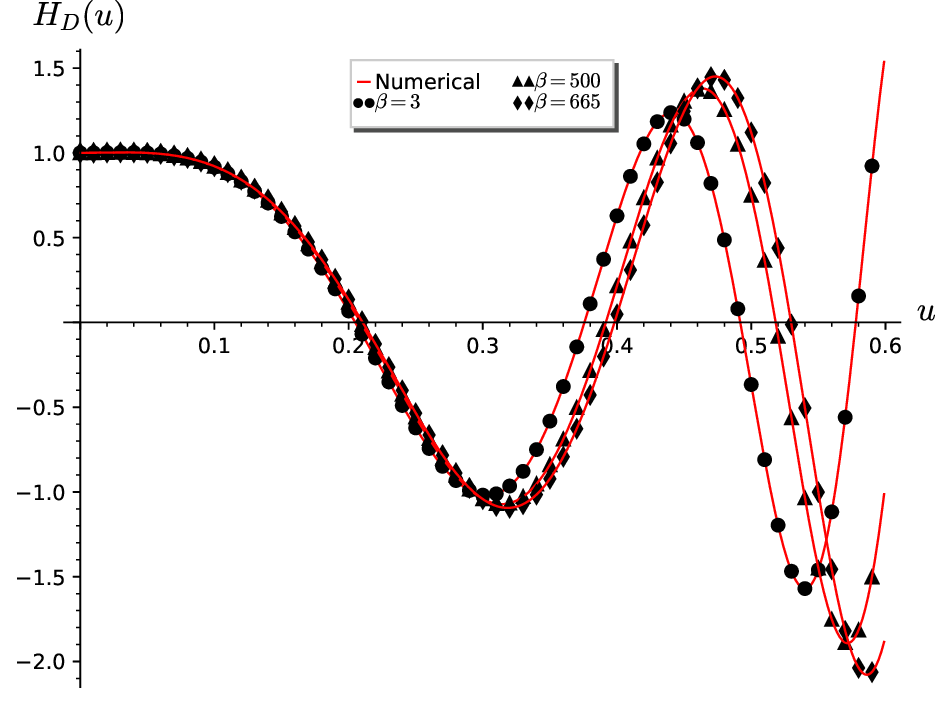}
		\caption{$H_D$ for varying $\beta$}
		\label{fig:DCH_beta}
	\end{subfigure}
	\begin{subfigure}{.5\textwidth}
		\centering
		% include second image
		\includegraphics[width=1.0\linewidth]{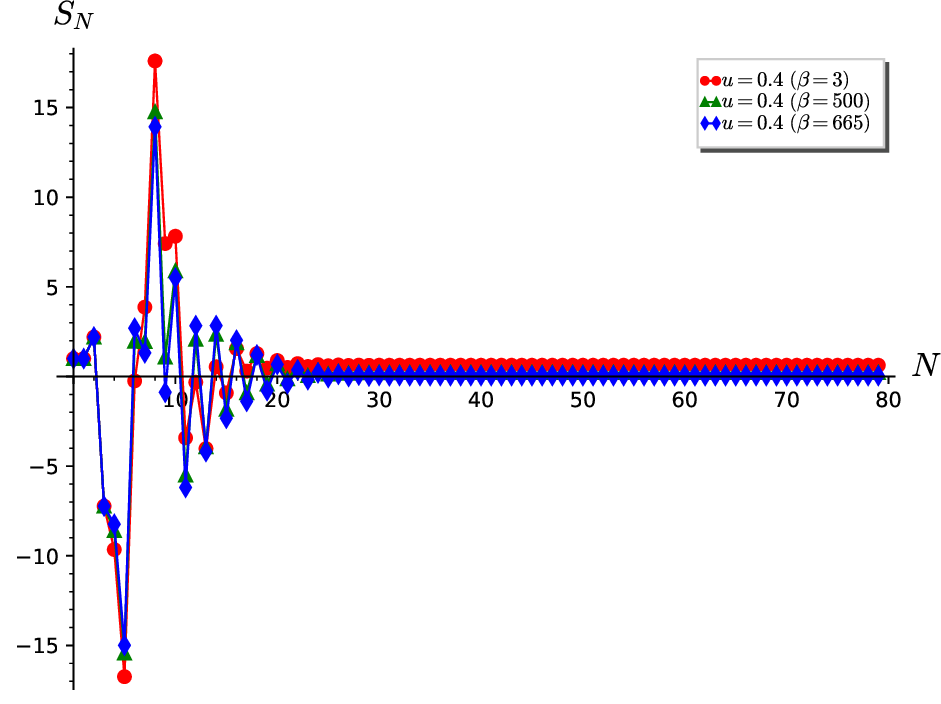}
		\caption{Convergence for $u=0.4$}
		\label{fig:DCH_conv_beta}
	\end{subfigure}
 \caption{ Analysis of the DCH series solution for varying $\beta$ with \{$ \alpha=3.0, \gamma=-935, \delta=15 $\}.} 
\end{figure}
As seen from the Figure (\ref{fig:DCH_conv_beta}), the series solutions converge to $S=0.63,0.22$ and $0.05$ for $\beta=3, \beta=500$ and $\beta=665$ at $u=0.4$, respectively.
\begin{figure}[H]
	\begin{subfigure}{.5\textwidth}
		\centering
		% include first image
		\includegraphics[width=1.0\linewidth]{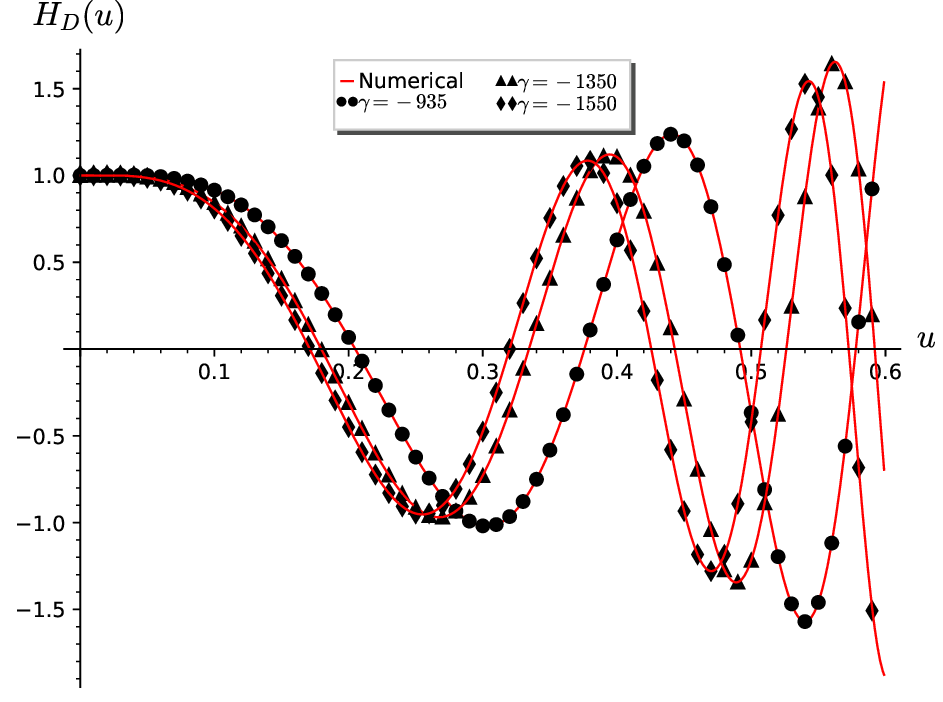}
		\caption{$H_D$ for varying $\gamma$}
		\label{fig:DCH_gamma}
	\end{subfigure}
	\begin{subfigure}{.5\textwidth}
		\centering
		% include second image
		\includegraphics[width=1.0\linewidth]{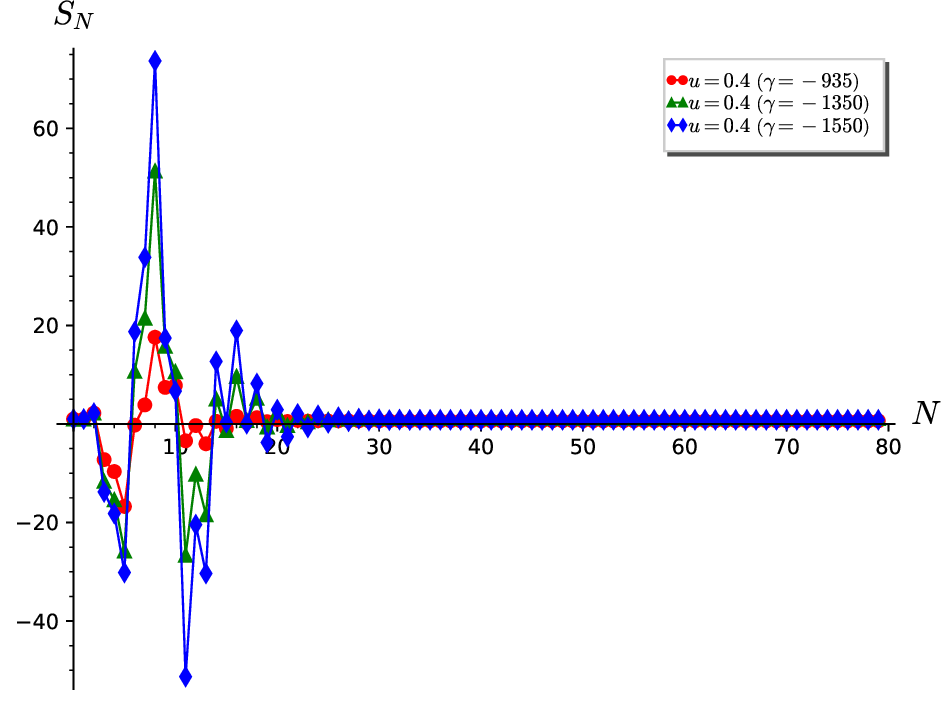}
		\caption{Convergence for $u=0.4$}
		\label{fig:DCH_conv_gamma}
	\end{subfigure}
\caption{ Analysis of the DCH series solution for varying $\gamma$ with \{$ \alpha=3, \beta=3, \delta=15 $\}.} 
\end{figure}
As seen from the Figure (\ref{fig:DCH_conv_gamma}), the series solutions converge to $S=0.63,1.11$ and $0.84$ for $\gamma=-935, \gamma=-1350$ and $\gamma=-1550$ at $u=0.4$, respectively.
\begin{figure}[H]
	\begin{subfigure}{.5\textwidth}
		\centering
		% include first image
		\includegraphics[width=1.0\linewidth]{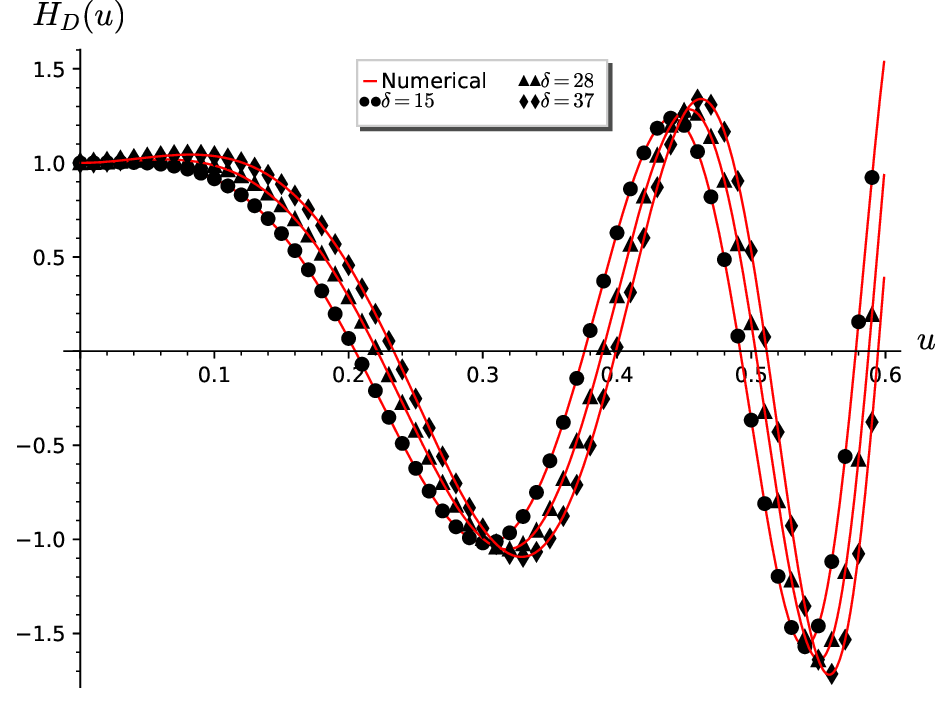}
		\caption{$H_D$ for varying $\delta$}
		\label{fig:DCH_delta}
	\end{subfigure}
	\begin{subfigure}{.5\textwidth}
		\centering
		% include second image
		\includegraphics[width=1.0\linewidth]{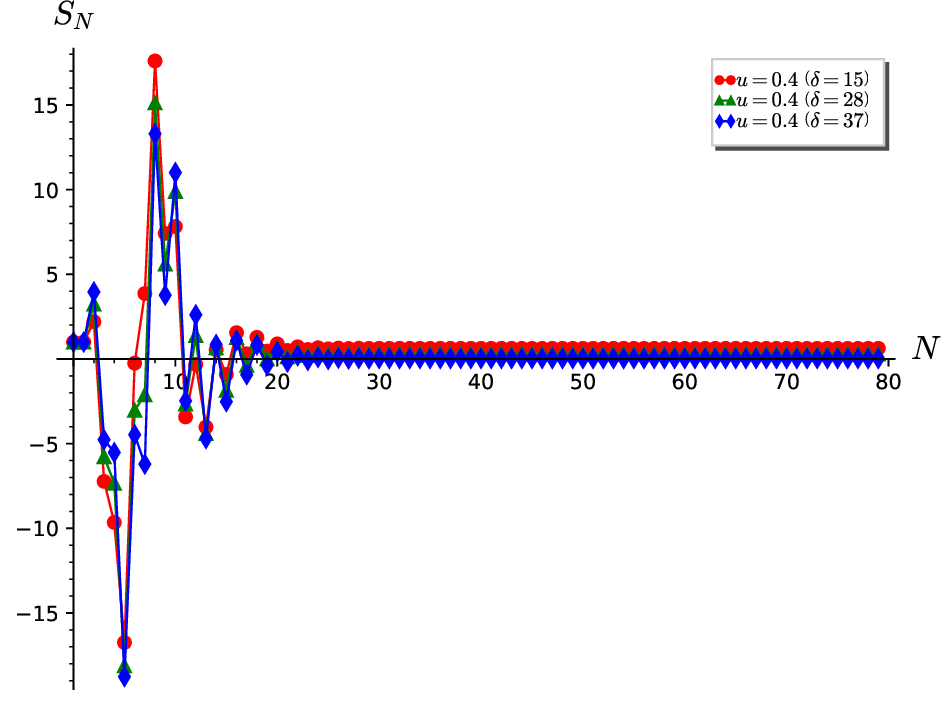}
		\caption{Convergence for $u=0.4$}
		\label{fig:DCH_conv_delta}
	\end{subfigure}
 \caption{ Analysis of the DCH series solution for varying $\delta$ with \{$ \alpha=3, \beta=3, \gamma=-935 $\}.}
\end{figure}
As seen from the Figure (\ref{fig:DCH_conv_delta}), the series solutions converge to $S=0.63,0.29$ and $0.02$ for $\delta=15, \delta=28$ and $\delta=37$ at $u=0.4$, respectively.
%%%%%%%%%%%%%%%%%%%%%%%%%%%%%%%%%%%%%%%%%%%%%%%%%%%%%%%%%%%%%%%%%%%%%%%%%%%%%%%%%%%%%
\section*{Appendix C: Series Solution Analysis of the DDCH Equation for Various Parameters}
\begin{figure}[H]
	\begin{subfigure}{.5\textwidth}
		\centering
		% include first image
		\includegraphics[width=1.0\linewidth]{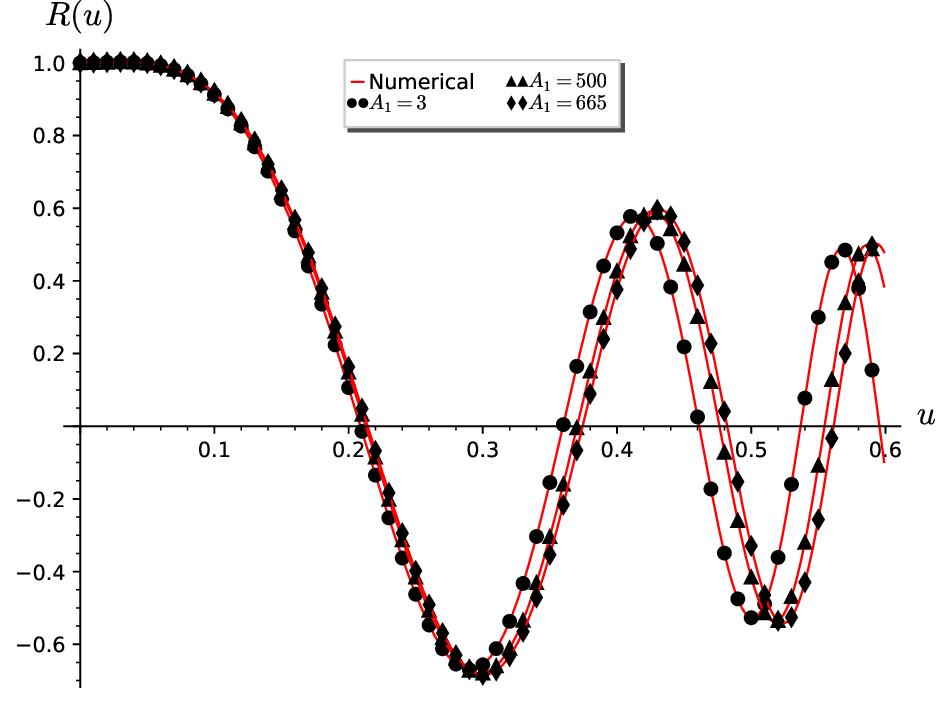}
		\caption{$R(u)$ for varying $A_1$}
		\label{fig:DDCH_A1}
	\end{subfigure}
	\begin{subfigure}{.5\textwidth}
		\centering
		% include second image
		\includegraphics[width=1.0\linewidth]{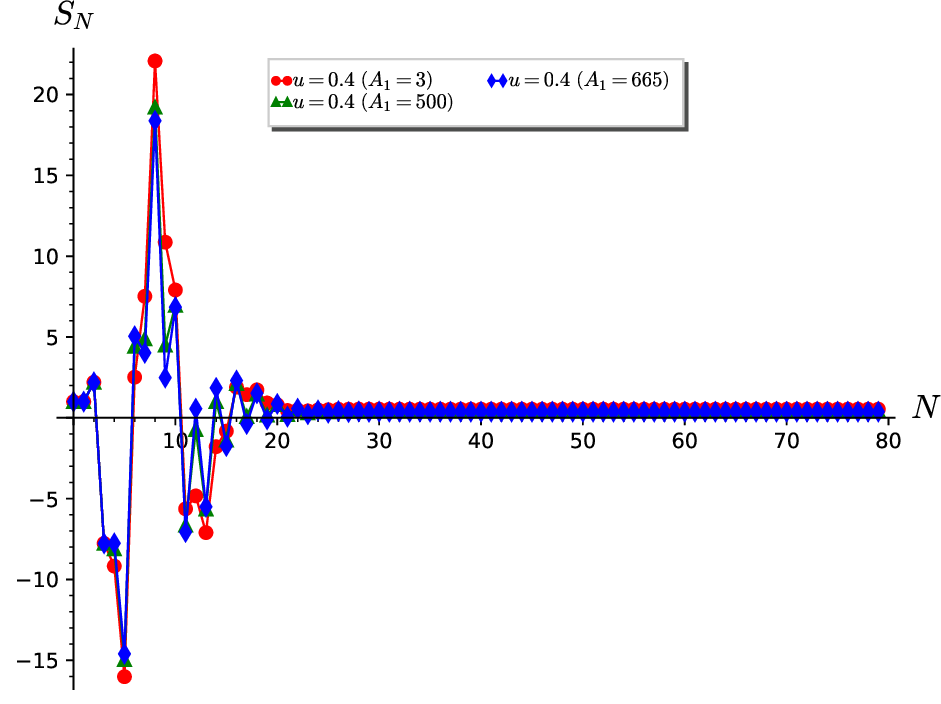}
		\caption{Convergence for $u=0.4$}
		\label{fig:DDCH_conv_A1}
	\end{subfigure}
\caption{ Analysis of the DDCH series solution for varying $A_1$  with \{$ A_2=\gamma=-935, A_3=\delta=15, A_4=-57, A_5=-278, A_6=-847$\}.} 
\end{figure}
As seen from the Figure (\ref{fig:DDCH_conv_A1}), the series solutions converge to $S=0.53,0.43$ and $0.38$ for $A_1=3, A_1=500$ and $A_1=665$ at $u=0.4$, respectively.
\begin{figure}[H]
	\begin{subfigure}{.5\textwidth}
		\centering
		% include first image
		\includegraphics[width=1.0\linewidth]{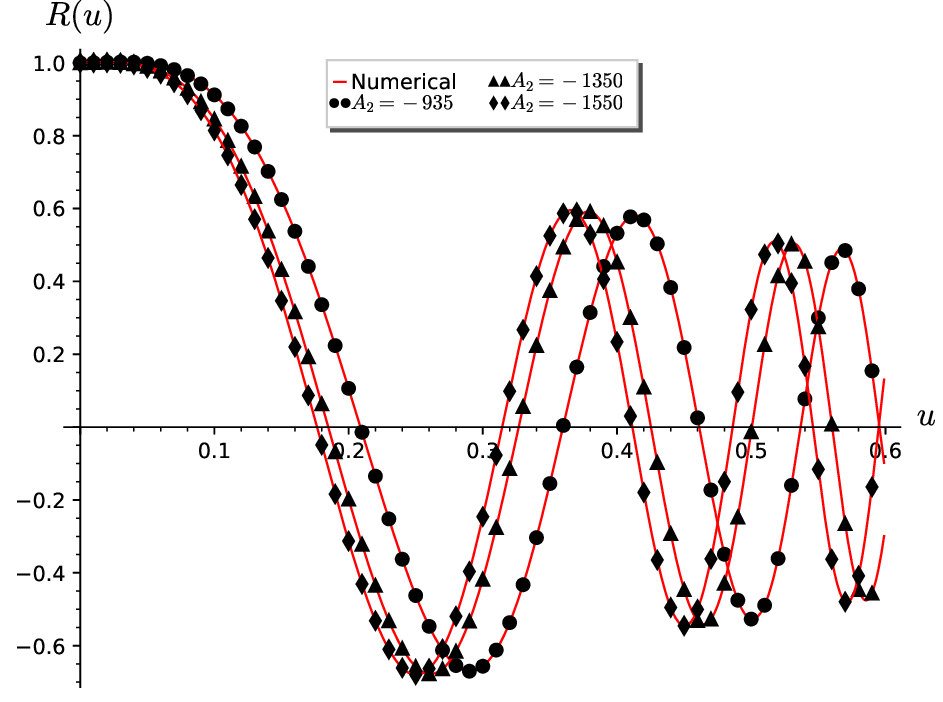}
		\caption{$R(u)$ for varying $A_2$}
		\label{fig:DDCH_A2}
	\end{subfigure}
	\begin{subfigure}{.5\textwidth}
		\centering
		% include second image
		\includegraphics[width=1.0\linewidth]{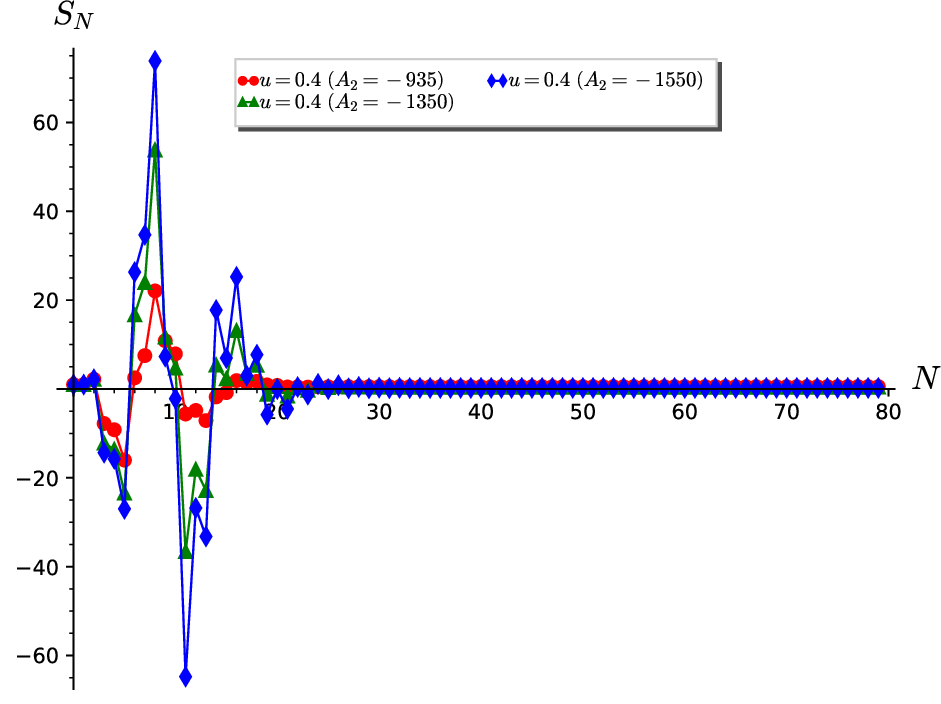}
		\caption{Convergence for $u=0.4$}
		\label{fig:DDCH_conv_A2}
	\end{subfigure}
\caption{ Analysis of the DDCH series solution for varying $A_2$  with \{$ A_1=\beta=3, A_3=\delta=15, A_4=-57, A_5=-278, A_6=-847$\}.} 
\end{figure}
As seen from the Figure (\ref{fig:DDCH_conv_A2}), the series solutions converge to $S=0.54,0.45$ and $0.23$ for $A_2=-935, A_2=-1350$ and $A_2=-1550$ at $u=0.4$, respectively.
\begin{figure}[H]
	\begin{subfigure}{.5\textwidth}
		\centering
		% include first image
		\includegraphics[width=1.0\linewidth]{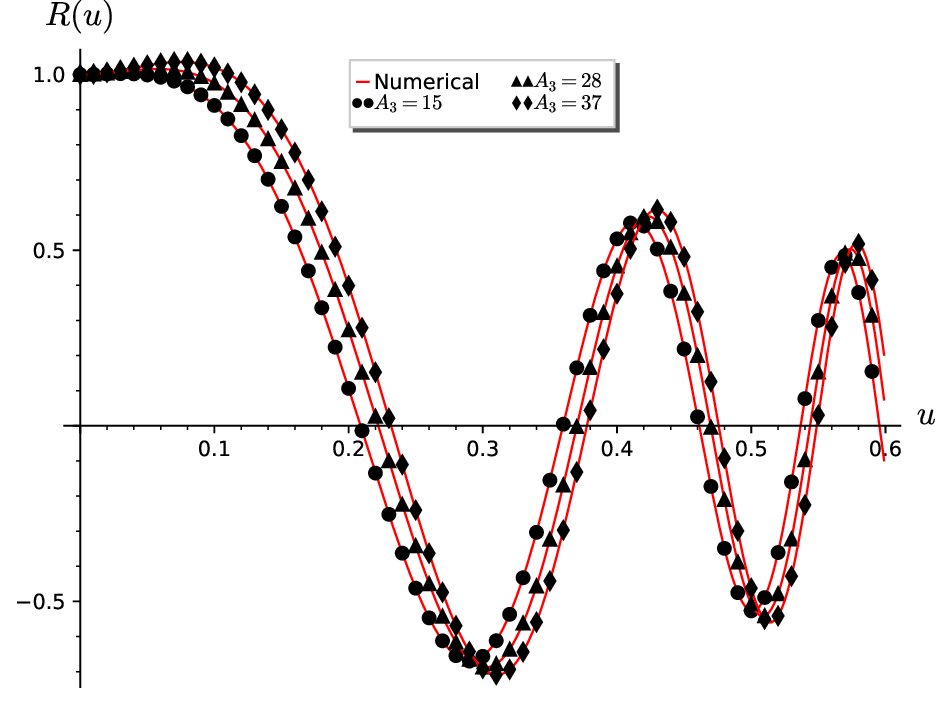}
		\caption{$R(u)$ for varying $A_3$}
		\label{fig:DDCH_A3}
	\end{subfigure}
	\begin{subfigure}{.5\textwidth}
		\centering
		% include second image
		\includegraphics[width=1.0\linewidth]{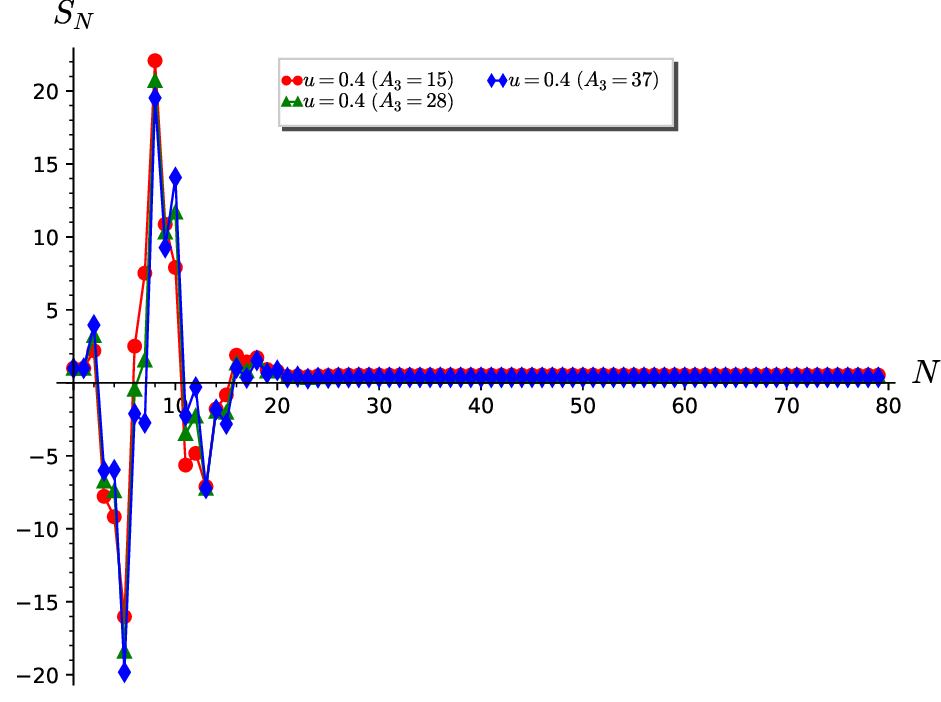}
		\caption{Convergence for $u=0.4$}
		\label{fig:DDCH_conv_A3}
	\end{subfigure}
\caption{ Analysis of the DDCH series solution for varying $A_3$  with \{$  A_1=\beta=3, A_2=\gamma=-935, A_4=-57, A_5=-278, A_6=-847$\}.} 
\end{figure}
As seen from the Figure (\ref{fig:DDCH_conv_A3}), the series solutions converge to $S=0.53,0.45$ and $0.38$ for $A_3=15, A_3=28$ and $A_3=37$ at $u=0.4$, respectively.
\begin{figure}[H]
	\begin{subfigure}{.5\textwidth}
		\centering
		% include first image
		\includegraphics[width=1.0\linewidth]{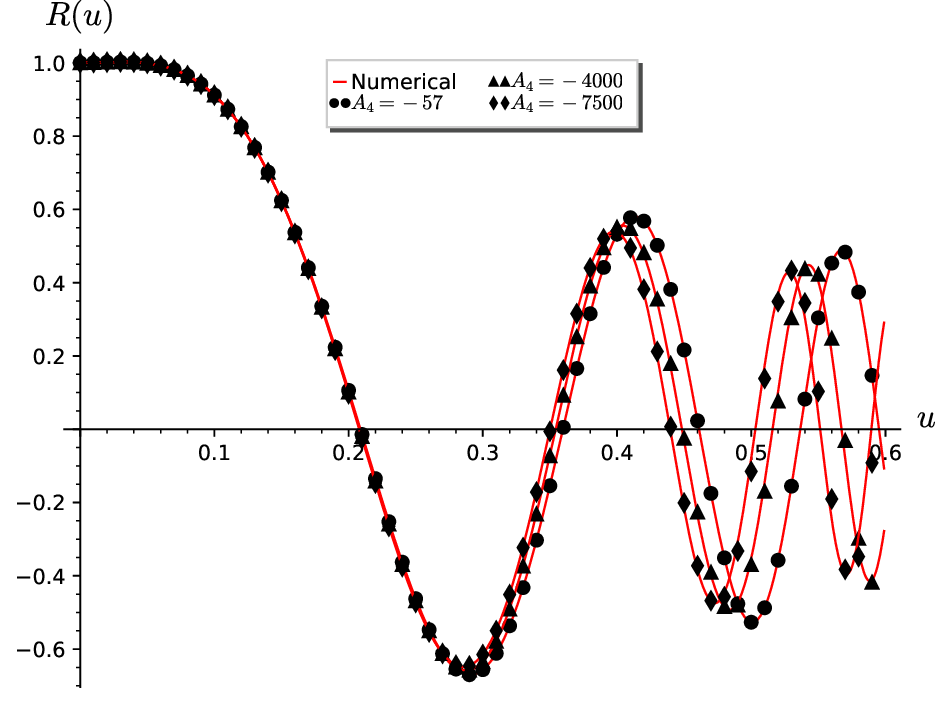}
		\caption{$R(u)$ for varying $A_4$}
		\label{fig:DDCH_A4}
	\end{subfigure}
	\begin{subfigure}{.5\textwidth}
		\centering
		% include second image
		\includegraphics[width=1.0\linewidth]{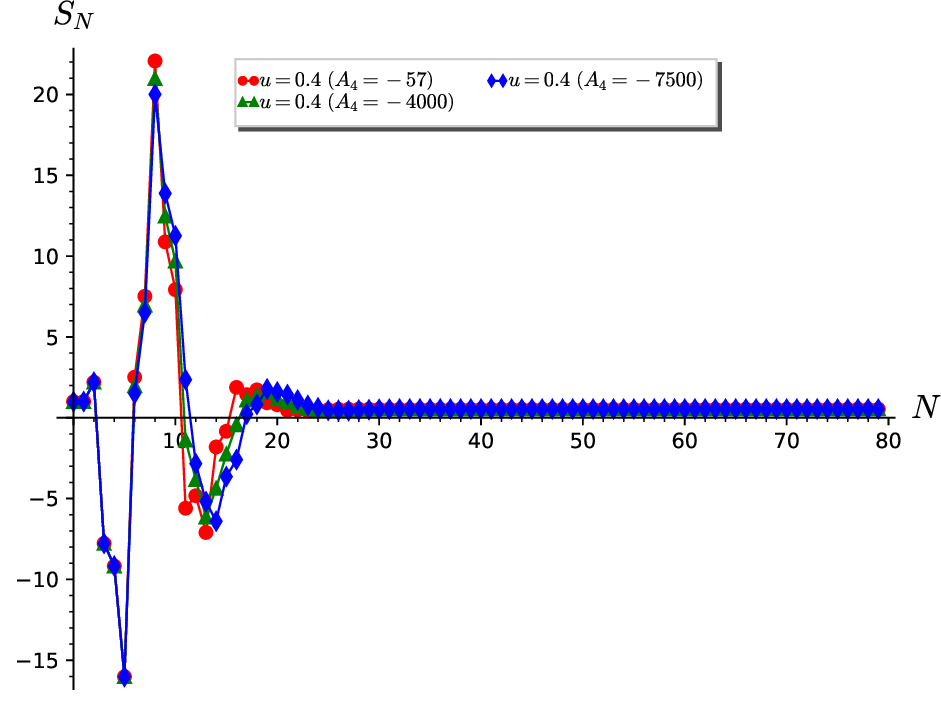}
		\caption{Convergence for $u=0.4$}
		\label{fig:DDCH_conv_A4}
	\end{subfigure}
\caption{ Analysis of the DDCH series solution for varying $A_4$  with \{$  A_1=\beta=3, A_2=\gamma=-935, A_3=15, A_5=-278, A_6=-847$\}.}
\end{figure}
As seen from the Figure (\ref{fig:DDCH_conv_A4}), the series solutions converge to $S=0.53,0.55$ and $0.54$ for $A_4=-57, A_4=-4000$ and $A_4=-7500$ at $u=0.4$, respectively.
\begin{figure}[H]
	\begin{subfigure}{.5\textwidth}
		\centering
		% include first image
		\includegraphics[width=1.0\linewidth]{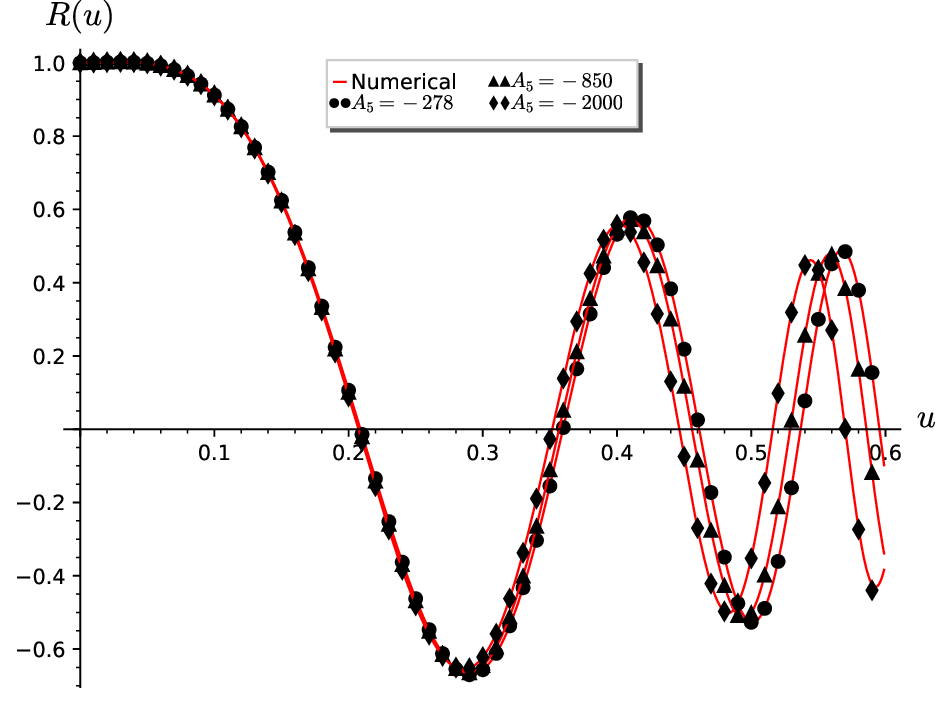}
		\caption{$R(u)$ for varying $A_5$}
		\label{fig:DDCH_A5}
	\end{subfigure}
	\begin{subfigure}{.5\textwidth}
		\centering
		% include second image
		\includegraphics[width=1.0\linewidth]{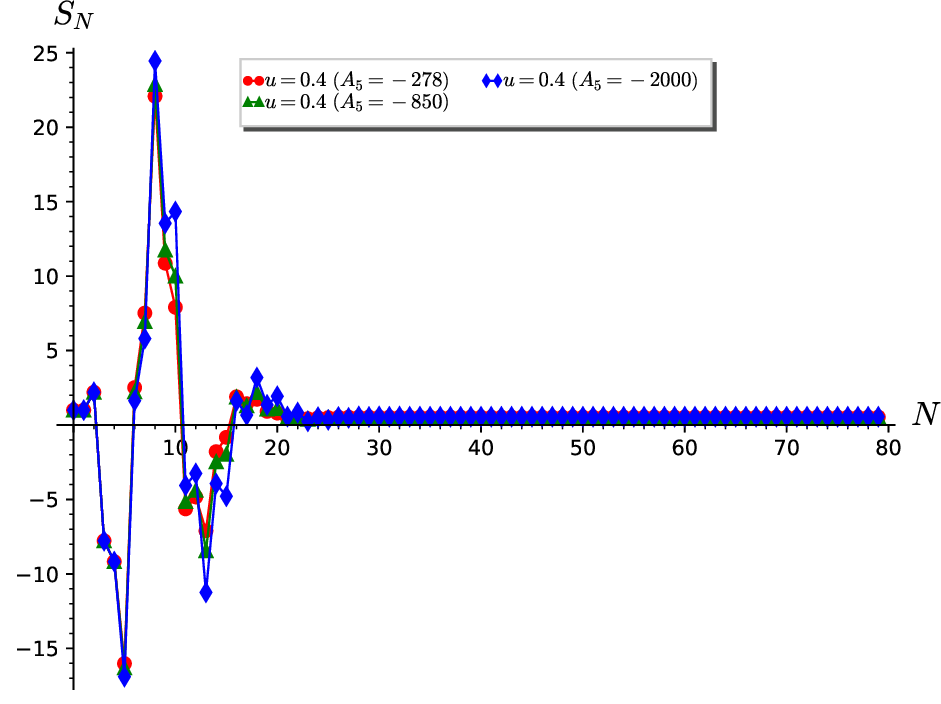}
		\caption{Convergence for $u=0.4$}
		\label{fig:DDCH_conv_A5}
	\end{subfigure}
\caption{ Analysis of the DDCH series solution for varying $A_5$  with \{$  A_1=\beta=3, A_2=\gamma=-935, A_3=15, A_4=-57, A_6=-847$\}.}
\end{figure}
As seen from the Figure (\ref{fig:DDCH_conv_A5}), the series solutions converge to $S=0.53,0.55$ and $0.56$ for $A_5=-278, A_5=-850$ and $A_5=-2000$ at $u=0.4$, respectively.
\begin{figure}[H]
	\begin{subfigure}{.5\textwidth}
		\centering
		% include first image
		\includegraphics[width=1.0\linewidth]{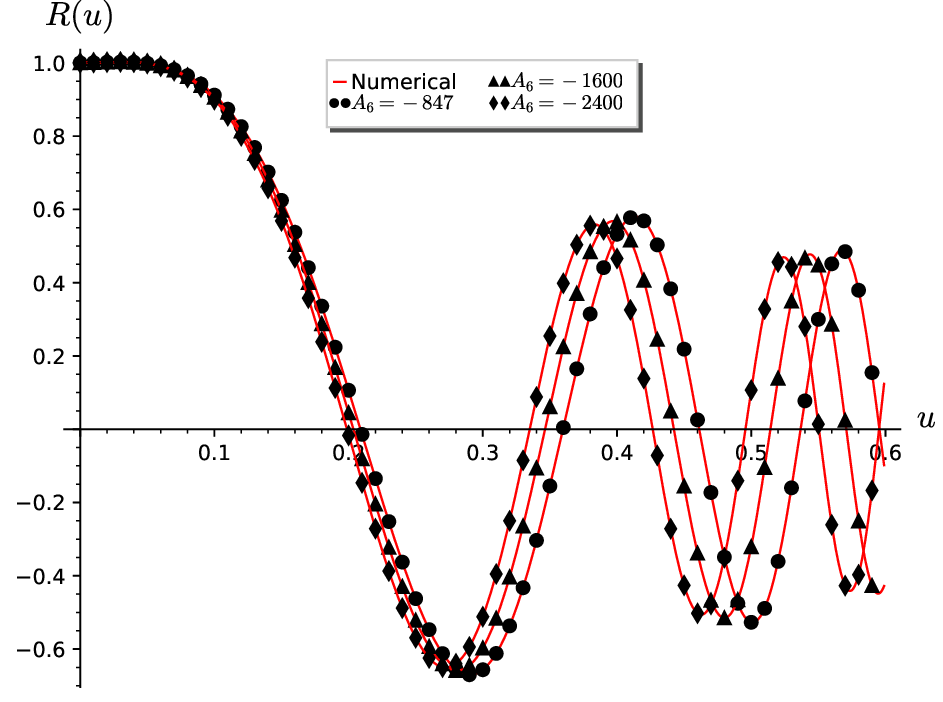}
		\caption{$R(u)$ for varying $A_6$}
		\label{fig:DDCH_A6}
	\end{subfigure}
	\begin{subfigure}{.5\textwidth}
		\centering
		% include second image
		\includegraphics[width=1.0\linewidth]{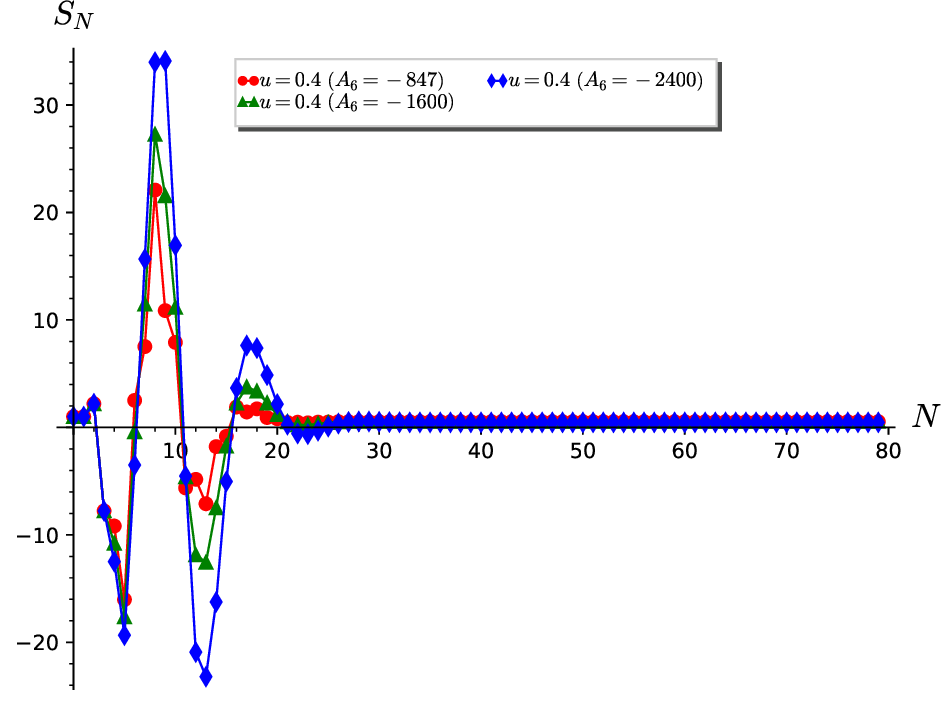}
		\caption{Convergence for $u=0.4$}
		\label{fig:DDCH_conv_A6}
	\end{subfigure}
\caption{ Analysis of the DDCH series solution for varying $A_6$  with \{$  A_1=\beta=3, A_2=\gamma=-935, A_3=15, A_4=-57, A_5=-278$\}.}\end{figure}
As seen from the Figure (\ref{fig:DDCH_conv_A6}), the series solutions converge to $S=0.53,0.57$ and $0.47$ for $A_6=-847, A_6=-1600$ and $A_6=-2400$ at $u=0.4$, respectively.
%%%%%%%%%%%%%%%%%%%%%%%%%%%%%%%%%%%%%%%%%%%%%
%%%%%%%%%%%%%%%%%%%%%%%%%%%%%%%%%%%%%%%%%%%%

\end{document}